\documentclass[aps,prd,superscriptaddress,floatfix,showpacs,
               twocolumn, 
]{revtex4-1}

\usepackage{lineno}
\usepackage{graphicx}
\usepackage{braket}
\usepackage{natbib} 
\usepackage{xspace}	
\usepackage{longtable}

\def\bslash#1{\setbox0=\hbox{$#1$}                     
    \dimen0=\wd0                                       
    \setbox1=\hbox{$\backslash$} \dimen1=\wd1          
    \ifdim\dimen0>\dimen1                              
       \rlap{\hbox to \dimen0{\hfil$\backslash$\hfil}} 
       #1                                              
    \else                                              
       \rlap{\hbox to \dimen1{\hfil$#1$\hfil}}         
       \backslash                                      
    \fi}                                               %

\newcommand{\cc}{\ensuremath{\textrm{CC}}\xspace}
\newcommand{\nc}{\ensuremath{\textrm{NC}}\xspace}
\newcommand{\piz}{\ensuremath{\pi^0}\xspace}
\newcommand{\pip}{\ensuremath{\pi^+}\xspace}
\newcommand{\pim}{\ensuremath{\pi^-}\xspace}
\newcommand{\mum}{\ensuremath{\mu^-}\xspace}
\newcommand{\kp}{\ensuremath{K^+}\xspace}
\newcommand{\kz}{\ensuremath{K^0}\xspace}

\newcommand{\ccpiz}{\ensuremath{\cc\piz}\xspace}
\newcommand{\ncpiz}{\ensuremath{\nc\piz}\xspace}
\newcommand{\ccpip}{\ensuremath{\cc\pip}\xspace}
\newcommand{\ncpip}{\ensuremath{\nc\pip}\xspace}

\newcommand{\pipabs}{\ensuremath{\pip\to\bslash{\pi}}\xspace}
\newcommand{\pipcex}{\ensuremath{\pip\to\piz}\xspace}
\newcommand{\piabs}{\ensuremath{\pi\to\bslash{\pi}}\xspace}

\newcommand{\numu}{\ensuremath{\nu_\mu}\xspace}
\newcommand{\nue}{\ensuremath{\nu_e}\xspace}
\newcommand{\anumu}{\ensuremath{\bar{\nu}_\mu}\xspace}
\newcommand{\anue}{\ensuremath{\bar{\nu}_e}\xspace}

\newcommand{\chtwo}{\ensuremath{\textrm{CH}_2}\xspace}

\newcommand{\pz}{\ensuremath{\phantom{0}}\xspace}

\newcommand{\che}{\v{C}erenkov\xspace}

\newcommand{\ccpizreac}{\ensuremath{\numu n\to\mum\piz p}\xspace}

\begin{document}


\title{Measurement of \numu-induced charged-current neutral pion
production cross sections on mineral oil at $E_\nu\in0.5-2.0$ GeV}

\newcommand{\bama}{University of Alabama; Tuscaloosa, AL 35487}
\newcommand{\bucknell}{Bucknell University; Lewisburg, PA 17837}
\newcommand{\cinci}{University of Cincinnati; Cincinnati, OH 45221}
\newcommand{\colorado}{University of Colorado; Boulder, CO 80309}
\newcommand{\columbia}{Columbia University; New York, NY 10027}
\newcommand{\embry}{Embry-Riddle Aeronautical University; Prescott, AZ 86301}
\newcommand{\fnal}{Fermi National Accelerator Laboratory; Batavia, IL 60510}
\newcommand{\florida}{University of Florida; Gainesville, FL 32611}
\newcommand{\indiana}{Indiana University; Bloomington, IN 47405}
\newcommand{\lanl}{Los Alamos National Laboratory; Los Alamos, NM 87545}
\newcommand{\lsu}{Louisiana State University; Baton Rouge, LA 70803}
\newcommand{\umich}{University of Michigan; Ann Arbor, MI 48109}
\newcommand{\princeton}{Princeton University; Princeton, NJ 08544}
\newcommand{\marys}{Saint Mary's University of Minnesota; Winona, MN 55987}
\newcommand{\vtech}{Virginia Polytechnic Institute \& State University; Blacksburg, VA 24061}
\newcommand{\yale}{Yale University; New Haven, CT 06520}
\newcommand{\beijing}{Institute of High Energy Physics; Beijing 100049, China}
\newcommand{\hope}{Hope College; Holland, MI 49423}
\newcommand{\iit}{Illinois Institute of Technology; Chicago, IL 60616}
\newcommand{\imsa}{Illinois Mathematics and Science Academy; Aurora IL 60506}
\newcommand{\massit}{Massachusetts Institute of Technology; Cambridge, MA 02139}
\newcommand{\caltech}{California Institute of Technology; Pasadena, CA 91125}
\newcommand{\bu}{Boston University; Boston, MA 02215}
\newcommand{\valencia}{IFIC, Universidad de Valencia and CSIC; 46071 Valencia, Spain}
\newcommand{\ubc}{University of British Columbia, Vancouver, BC V6T 1Z1, Canada}
\newcommand{\imperial}{Imperial College; London SW7 2AZ, United Kingdom}
\newcommand{\mexi}{Instituto de Ciencias Nucleares, Universidad Nacional Aut\'onoma de M\'exico, D.F. 04510, M\'exico}
\newcommand{\argonne}{Argonne National Laboratory; Argonne, IL 60439}

\affiliation{\bama}
\affiliation{\argonne}
\affiliation{\bucknell}
\affiliation{\cinci}
\affiliation{\colorado}
\affiliation{\columbia}
\affiliation{\embry}
\affiliation{\fnal}
\affiliation{\florida}
\affiliation{\indiana}
\affiliation{\lanl}
\affiliation{\lsu}
\affiliation{\massit}
\affiliation{\mexi}
\affiliation{\umich}
\affiliation{\princeton}
\affiliation{\marys}
\affiliation{\vtech}
\affiliation{\yale}

\author{A.~A. Aguilar-Arevalo}\affiliation{\mexi}
\author{C.~E.~Anderson}\affiliation{\yale}
\author{A.~O.~Bazarko}\affiliation{\princeton}
\author{S.~J.~Brice}\affiliation{\fnal}
\author{B.~C.~Brown}\affiliation{\fnal}
\author{L.~Bugel}\affiliation{\massit}
\author{J.~Cao}\affiliation{\umich}
\author{L.~Coney}\affiliation{\columbia}
\author{J.~M.~Conrad}\affiliation{\massit}
\author{D.~C.~Cox}\affiliation{\indiana}
\author{A.~Curioni}\affiliation{\yale}
\author{R.~Dharmapalan}\affiliation{\bama}
\author{Z.~Djurcic}\affiliation{\argonne}
\author{D.~A.~Finley}\affiliation{\fnal}
\author{B.~T.~Fleming}\affiliation{\yale}
\author{R.~Ford}\affiliation{\fnal}
\author{F.~G.~Garcia}\affiliation{\fnal}
\author{G.~T.~Garvey}\affiliation{\lanl}
\author{J.~Grange}\affiliation{\florida}
\author{C.~Green}\affiliation{\fnal}\affiliation{\lanl}
\author{J.~A.~Green}\affiliation{\indiana}\affiliation{\lanl}
\author{T.~L.~Hart}\affiliation{\colorado}
\author{E.~Hawker}\affiliation{\cinci}\affiliation{\lanl}
\author{R.~Imlay}\affiliation{\lsu}
\author{R.~A.~Johnson}\affiliation{\cinci}
\author{G.~Karagiorgi}\affiliation{\massit}
\author{P.~Kasper}\affiliation{\fnal}
\author{T.~Katori}\affiliation{\indiana}\affiliation{\massit}
\author{T.~Kobilarcik}\affiliation{\fnal}
\author{I.~Kourbanis}\affiliation{\fnal}
\author{S.~Koutsoliotas}\affiliation{\bucknell}
\author{E.~M.~Laird}\affiliation{\princeton}
\author{S.~K.~Linden}\affiliation{\yale}
\author{J.~M.~Link}\affiliation{\vtech}
\author{Y.~Liu}\affiliation{\umich}
\author{Y.~Liu}\affiliation{\bama}
\author{W.~C.~Louis}\affiliation{\lanl}
\author{K.~B.~M.~Mahn}\affiliation{\columbia}
\author{W.~Marsh}\affiliation{\fnal}
\author{C.~Mauger}\affiliation{\lanl}
\author{V.~T.~McGary}\affiliation{\massit}
\author{G.~McGregor}\affiliation{\lanl}
\author{W.~Metcalf}\affiliation{\lsu}
\author{P.~D.~Meyers}\affiliation{\princeton}
\author{F.~Mills}\affiliation{\fnal}
\author{G.~B.~Mills}\affiliation{\lanl}
\author{J.~Monroe}\affiliation{\columbia}
\author{C.~D.~Moore}\affiliation{\fnal}
\author{J.~Mousseau}\affiliation{\florida}
\author{R.~H.~Nelson}\altaffiliation{Present address: \caltech}\affiliation{\colorado}
\author{P.~Nienaber}\affiliation{\marys}
\author{J.~A.~Nowak}\affiliation{\lsu}
\author{B.~Osmanov}\affiliation{\florida}
\author{S.~Ouedraogo}\affiliation{\lsu}
\author{R.~B.~Patterson}\affiliation{\princeton}
\author{Z.~Pavlovic}\affiliation{\lanl}
\author{D.~Perevalov}\affiliation{\bama}\affiliation{\fnal}
\author{C.~C.~Polly}\affiliation{\fnal}
\author{E.~Prebys}\affiliation{\fnal}
\author{J.~L.~Raaf}\affiliation{\cinci}
\author{H.~Ray}\affiliation{\florida}
\author{B.~P.~Roe}\affiliation{\umich}
\author{A.~D.~Russell}\affiliation{\fnal}
\author{V.~Sandberg}\affiliation{\lanl}
\author{R.~Schirato}\affiliation{\lanl}
\author{D.~Schmitz}\affiliation{\fnal}
\author{M.~H.~Shaevitz}\affiliation{\columbia}
\author{F.~C.~Shoemaker}\altaffiliation{deceased}\affiliation{\princeton}
\author{D.~Smith}\affiliation{\embry}
\author{M.~Soderberg}\affiliation{\yale}
\author{M.~Sorel}\altaffiliation{Present address: \valencia}\affiliation{\columbia}
\author{P.~Spentzouris}\affiliation{\fnal}
\author{J.~Spitz}\affiliation{\yale}
\author{I.~Stancu}\affiliation{\bama}
\author{R.~J.~Stefanski}\affiliation{\fnal}
\author{M.~Sung}\affiliation{\lsu}
\author{H.~A.~Tanaka}\affiliation{\princeton}
\author{R.~Tayloe}\affiliation{\indiana}
\author{M.~Tzanov}\affiliation{\colorado}
\author{R.~G.~Van~de~Water}\affiliation{\lanl}
\author{M.~O.~Wascko}\altaffiliation{Present address: \imperial}\affiliation{\lsu}
\author{D.~H.~White}\affiliation{\lanl}
\author{M.~J.~Wilking}\affiliation{\colorado}
\author{H.~J.~Yang}\affiliation{\umich}
\author{G.~P.~Zeller}\affiliation{\fnal}
\author{E.~D.~Zimmerman}\affiliation{\colorado}

\collaboration{MiniBooNE Collaboration}
\noaffiliation

\date{\today}

\begin{abstract}
Using a custom 3 \che-ring fitter, 
we report cross sections for \numu-induced 
charged-current single \piz production on mineral oil (\chtwo) from a sample of 5810
candidate events with 57\% signal purity over an energy range of
$0.5-2.0$~GeV.  This includes measurements of the absolute total cross
section as a function of neutrino energy, and flux-averaged differential
cross sections measured in terms of $Q^2$, \mum kinematics, and \piz
kinematics.  The sample yields a flux-averaged total cross section of
$(9.2\pm0.3_{stat.}\pm1.5_{syst.})\times10^{-39}$~cm$^2$/\chtwo at
mean neutrino energy of $0.965$~GeV.   
\end{abstract}


\maketitle

\section{Introduction}
The charged-current interaction of a muon neutrino producing a single
neutral pion (\ccpiz) most commonly occurs through the $\Delta(1232)$
resonance for neutrino energies below 2~GeV.  As there is no coherent 
contribution to \ccpiz production, this process is an ideal probe of purely 
incoherent pion-production processes and thus offers additional kinematic
information on $\pi^0$ production beyond what is measured in the 
neutral-current channel~\cite{boone:ncpi0:2,sciboone:ncpi0}. Previous measurements 
of \ccpiz production at these energies were made on deuterium at the ANL 
12~ft bubble chamber~\cite{anl:barish,anl:radecky} and the BNL 7~ft bubble 
chamber~\cite{bnl:kitagaki}. Total cross-section measurements were reported 
on samples of 202.2~\cite{anl:radecky} and 853.5~\cite{bnl:kitagaki} events 
for the ANL and BNL experiments respectively. Previous 
measurements~\cite{cern:allasia,skat:grabosch} were also performed at higher
neutrino energy on a variety of targets.  

Using \che light detection techniques, this paper revisits this topic 
and measures \ccpiz production on carbon. 
In order to extract such interactions 
from the more dominant charged-current quasi-elastic (CCQE) and charged-current 
single \pip (\ccpip) production processes, a custom fitter has been developed to 
isolate and fit both the \mum and the \piz in a \ccpiz event. This fitter also accurately 
reconstructs the kinematics of these interactions providing a means with which to extract 
both total and single-differential cross sections. Additionally reported is a measurement 
of the flux-averaged total cross section. This work presents the most comprehensive 
measurements of \ccpiz interactions to date, at energies below 2~GeV, on a sample of 
events 3.5 times that of the combined previous measurements. Results include 
the total cross section, the single-differential cross section in $Q^2$, and the first 
measurements of single-differential cross sections in terms of final-state particle
kinematics. The reported cross sections provide a combined measure of the
primary interaction cross section, nuclear effects in carbon, and pion
re-interactions in the target nucleus.

\section{Final state interactions \& observable \ccpiz}
Because this measurement is being performed on a nuclear target, particular 
attention must be paid to how the sample is being defined, especially given
how nuclear and final-state effects can influence the observables. 
The dominant effect is final state interactions (FSI) which are the
re-interactions of particles created from the neutrino-nucleon interaction
with the nuclear medium of the target nucleus.  FSI change the
experimental signature of a neutrino-nucleon interaction.  For example, if a
\pip from a \ccpip interaction is absorbed
within the target nucleus and none of the outgoing nucleons are detected,
then the event is indistinguishable from a 
CCQE interaction. Additionally, if the \pip
charge exchanges then the interaction is indistinguishable from a \ccpiz
interaction.  This is due to the fact that the nuclear debris is typically 
unobservable in a \che-style detector.   
The understanding of FSI effects is model-dependent, with large
uncertainties on the FSI cross sections. An ``observable'' interaction is
therefore defined by the leptons and mesons that remain after FSI
effects.  Observable interactions are also inclusive of all nucleon final
states.  
To reduce the FSI-model dependence of the measurements reported here, the signal is
defined as a \mum and a single \piz that exits the target nucleus, with any
number of nucleons, and with
no additional mesons or leptons surviving the nucleus.  This is referred
to as an observable \ccpiz event.  The results presented here are not corrected
for nuclear effects and intra-nuclear interactions. 

\section{The MiniBooNE experiment}
The Mini Booster Neutrino Experiment (MiniBooNE)~\cite{boone:prop} is a
high-statistics low-energy neutrino experiment located at Fermilab.  A beam
of 8~GeV kinetic energy protons is taken from the
Booster~\cite{fnal:booster} and impinged upon a 71~cm long beryllium target.
The data set presented in this paper corresponds to
$6.27\times10^{20}$~p.o.t.~(protons on target) with an uncertainty of 2\%.  
The resulting pions and kaons are (de)focused according to their charge by a
toroidal magnetic field created by an aluminum magnetic focusing horn;
positive charge selection is used for this data.  These mesons then 
decay in a 50~m long air-filled pipe before the remnant beam impacts a steel beam dump.
The predominantly \numu-neutrino beam passes through 500~m of dirt
before interacting in a spherical 800~ton, 12~m diameter, mineral oil (\chtwo),
\che detector.  The center of the detector is positioned 541~m from
the beryllium target.  The inner surface of the detector is painted black
and instrumented with 1280 inward-facing 8~inch photomultiplier-tubes (PMTs)
providing 11.3\% photocathode coverage. A thin, optically-isolated shell
surrounds the main tank region and acts to veto entering and exiting charged particles
from the main tank. The veto region is painted white and instrumented with
240 tangentially-facing 8~inch PMTs.  A full description of the MiniBooNE
detector can be found in Ref.~\cite{boone:detnim}.

The neutrino beam is simulated within a {\sc Geant4}~\cite{geant4} Monte
Carlo (MC) framework.  All relevant components of the primary proton
beam line, beryllium target, aluminum focusing horn, collimator, meson decay
volume, beam dump, 
and surrounding earth are modeled~\cite{boone:flux}.  The total p-Be and
p-Al cross sections are set by the Glauber model~\cite{glauber}.  Wherever
possible, inelastic production cross sections are fit to external data.  The neutrino
beam is dominated by \numu produced by \pip decay in flight.  The \pip
production cross sections are set by a Sanford-Wang~\cite{sanwang} fit to
\pip production data provided by the HARP~\cite{harp} and E910~\cite{e910}
experiments.  The high-energy ($E_\nu>2.4$~GeV) neutrino flux is dominated
by \numu from \kp decays.  The \kp production cross sections are set by
fitting data from
Refs.~\cite{aleshin,abbott,allaby,dekkers,eichten,lundy,marmer,vorontsov} to
a Feynman scaling parametrization.  The production of protons and neutrons
on the target are set using the {\sc MARS}~\cite{mars} simulation.  The
\nue, \anumu, and \anue contributions to the flux are unimportant for this
measurement.  Ref.~\cite{boone:flux} describes the full details of the
neutrino-flux prediction and estimation of its systematic uncertainty.
It should be noted that the MiniBooNE neutrino data has not been used to
tune the flux prediction.

Interactions of neutrinos with the detector materials are simulated using
the v3 {\sc Nuance} event generator~\cite{nuance}.  The {\sc Nuance} event
generator is a comprehensive simulation of 99 neutrino and anti-neutrino
interactions on nuclear targets over an energy range from 100~MeV to 1~TeV.
The dominant interaction in MiniBooNE, CCQE, is modeled
according to Smith-Moniz~\cite{model:SM}; however, the axial mass, $M_A$, has been
adjusted for better agreement with the MiniBooNE data to 
$M_A^{\mathrm{eff}}=1.23\pm0.20$~GeV/$c^2$~\cite{boone:ccqe}.  
The target nucleus is simulated with nucleons bound in a relativistic
Fermi gas (RFG)~\cite{model:SM} with binding energy
$E_B=34\pm9$~MeV and Fermi momentum $p_F=220\pm30$~MeV/$c$~\cite{data:ES} (on
carbon).  The RFG model is further modified by shape fits to $Q^2$, for
better agreement of the CCQE interaction to MiniBooNE data~\cite{boone:ccqe}. 
The Rein-Sehgal
model~\cite{model:RS} is used to predict the production of single-pion final
states for both CC and NC modes.  This model includes 18 non-strange baryon
resonances below 2~GeV in mass and their interference terms.  
The model~\cite{model:RS} predicts the $\Delta(1232)$ resonance to account for
71\% of \ccpiz production (84\% of resonant production); a 14\% contribution
from higher mass resonances; a 15\% contribution from non-resonant processes.
The non-resonant processes are added {\em ad hoc} to improve agreement to
past data~\cite{model:RS}. They and are not indicative of actual non-resonant
contributions but allow for additional inelastic contributions.  The quarks  
are modeled as relativistic harmonic oscillators {\it a la} the 
Feynman-Kislinger-Ravndal model~\cite{model:FKR}.  The axial mass for single-pion
production is $M_A^{1\pi}=1.10\pm0.27$~GeV/$c^2$, and for multi-pion
production it is $M_A^{N\pi}=1.30\pm0.52$~GeV/$c^2$.   The model includes the
re-interactions of both baryon resonances, pions, and nucleons with the spectator
nucleons leading to the production of additional pions, pion
charge-exchange, and pion absorption.  For observable \ccpip interactions,
the model is reweighted to match the MiniBooNE data by a technique described
in \S\ref{s:ccpip}.

The MiniBooNE detector is simulated within a {\sc Geant3}~\cite{geant3}
MC.  This MC handles the propagation of particles after they exit the
neutrino-target nucleus, subsequent interactions with the mineral oil~\cite{calor}, and
most importantly, the propagation and interactions of optical photons.  
Photons with wavelengths between 250-650~nm are considered.  The production,
scattering, fluorescence, absorption, and reflections of these photons are
modeled in a 35 parameter custom optical
model~\cite{boone:detnim,thesis:ryan,brown:2004uy}.   
The detector MC also simulates photon detection by the PMTs and the effects
of detector electronics.  The absorption (\pipabs) and charge-exchange
(\pipcex) of \pip particles on carbon are fixed to external
data~\cite{exp:ashery,exp:piabscex:1,exp:piabscex:2} with uncertainties of
35\% and 50\% respectively. 

\section{Event reconstruction}\label{s:reco}
Particles traversing the mineral oil are detected by the \che and
scintillation light they produce.  The relative abundances of these emissions,
along with the shape of the total \che angular distribution, are used to classify the
type of particle in the detector.   For a single particle, an ``extended
track'' is fit using a maximum-likelihood method for several
possible particle hypotheses.  For each considered particle type, the
likelihood is a function of the initial vertex, kinetic energy, and
direction.  For a given set of track parameters, the likelihood function
calculates probability density functions (PDFs) for each of the 1280 PMTs in
the main portion of the detector~\cite{boone:recnim,thesis:ryan}. Separate
PDFs for the initial hit time and total integrated charge are produced.  As
the data acquisition records only the initial hit time and total charge for
each PMT, the \che and scintillation contributions are indistinguishable for
a given hit; however, they are distinguishable statistically.
The likelihood is formed as the product of the probabilities, calculated
from the PDFs. 
The initial track parameters are varied using {\sc
Minuit}~\cite{minuit} and the results of the best fit likelihood determines
the parameters for both the particle type and kinematics.   

The extended-track reconstruction is scalable to any number of tracks.  The charge
PDFs are constructed by adding the predicted charges from each track together to
determine the overall predicted charge.  The time PDFs are
calculated separately for each track, and separately for the \che and
scintillation portions, then time sorted and weighted by the 
probability that a particular PDF caused the initial
hit~\cite{boone:detnim,thesis:ryan}.  The reconstruction needed for 
an observable \ccpiz event requires three tracks: a \mum track and two photon
tracks from a common 
vertex~\cite{thesis:rhn,nuint09:rhn}.  The final state is defined by a \mum,
a \piz, and nuclear debris.  The \mum is directly fit by the reconstruction, along with
its decay electron.  The \piz decays into two photons with a branching
fraction of 98.8\%~\cite{pdg} at the neutrino interaction vertex
($c\tau=25.1$~nm).  The two photons are fit by the reconstruction.  Photons
produce \che rings both by converting ($\lambda=67$~cm) into $e^+e^-$ pairs
through interactions with the mineral oil and by Compton scattering.  The
nuclear debris is ignored in 
the reconstruction as it is rarely above \che threshold and therefore
only contributes to scintillation light. As the calculation of the kinetic
energy of a track is dominated by the \che ring, the added scintillation
light is effectively split uniformly among the three tracks and justifiably
ignored.   

The novelty of the \ccpiz event reconstruction is the ability to find and
reconstruct three \che rings.  The \ccpiz likelihood function is
parametrized by the event vertex ($x,y,z$), the event start time ($t$), the \mum
direction and kinetic energy ($\theta_\mu,\phi_\mu,E_\mu$), the first
$\gamma$ direction and energy ($\theta_1,\phi_1,E_1$), the first
$\gamma$ conversion length ($s_1$), the second 
$\gamma$ direction and energy ($\theta_2,\phi_2,E_2$), and conversion length
($s_2$).  Ring, or track finding is performed in a
stepwise fashion.  The first track is seeded in the likelihood function and
fixed by the $\mu$ one-track fit described in Ref.~\cite{boone:recnim}.
The second track is scanned through 400 evenly-spaced points in solid angle
assuming 200~MeV kinetic energy and no $\gamma$ conversion length.  The best
scan point 
is then allowed to float for both tracks simultaneously.  The third track is
found by fixing the two tracks and scanning again in solid angle for the
third track.  Once the third track is found, the best-scan point, along with
the two fixed tracks, are allowed to float.  This stage of the fit, referred
to as the ``generic'' three-track fit, determines a seed for the event vertex,
track directions, and track energies.  A series of three parallel fits are
then 
performed to determine track particle types and the final fit kinematics.
Each of these fits is seeded with the generic three-track fit.  The fits
assign a \mum hypothesis to one of the tracks and a $\gamma$ hypothesis to the
other two, allowing for the possibility that the \mum was not found by the
original $\mu$ one-track fit and was found during the first or second
scan.  The conversion lengths are seeded at 50~cm and fit 
along with the kinetic energies while keeping all the other parameters
fixed, thereby determining their seeds for the final portion of the fit.  The
final stage of each fit allows all parameters to float, taking advantage of
{\sc Minuit}'s {\sc Improve} function~\cite{minuit}.  A term is added to the
fit negative-log-likelihoods comparing the direction from the fit event
vertex to the fit \mum decay vertex versus the fit \mum direction weighting by
the separation of the vertices.  This term improves the
identification of the particle types.   The likelihoods are then compared to
choose the best fit.  For further details see Ref.~\cite{thesis:rhn}.

\begin{figure}[ht!]
\includegraphics[scale=0.4]{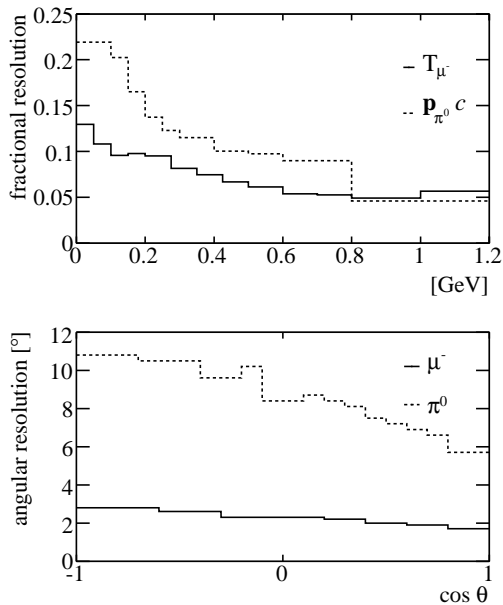}
\caption{Top: Fractional Gaussian resolutions as functions of $T_\mu$ (solid) and
$|\mathbf{p}_{\piz}|c$ (dashed).  Bottom: Angular resolutions as functions of
$\cos\theta_\mu$ (solid) and $\cos\theta_{\piz}$ (dashed).  The resolutions
between each particle type are also correlated.} 
\label{fig:reco}
\end{figure}
The quality of the reconstruction is assessed by evaluating the residual
resolutions of the signal using the MC.  Fig.~\ref{fig:reco} shows the
residual resolutions for both the \mum and the \piz.  The overall \mum
kinetic-energy fractional resolution is 7.4\%, and the angular resolution is
$2^\circ$.  The \piz, being a combination of the two fit photons, has an 
overall momentum resolution of 12.5\% and angular resolution of $7.8^\circ$.
The energy and momentum resolutions are worse at lower energies and momenta and flatten out
toward larger values.  The \mum angular resolution is mostly flat and is
only slightly better in the forward direction.  The \piz angular resolution
gets much better in the forward direction as forward going \piz tend to have
larger momentum.  The resolutions between each particle type are also
somewhat correlated.  Additionally, the interaction-vertex resolution is 16 cm. 

The initial neutrino energy is calculated, assuming that the signal events
are from the reaction \ccpizreac, from
the measured \mum and \piz kinematics under three assumptions: the
interaction target is a stationary neutron, the hadronic recoil is a proton,
and the neutrino is traveling in the beam direction.  Under these
assumptions, $E_\nu$ is constrained even if the proton is
unmeasured~\cite{nuint09:rhn}.   The assumption of a stationary neutron
contributes to smearing of the reconstructed neutrino energy because of the
neutron's momentum distribution.  The neutrino-energy resolution is 11\%.
Ideally one would measure the proton, and additional hadronic debris;
however, as these particles are rarely above \che threshold, it is impractical to
do so in MiniBooNE. Also, as only 70\% of the observable \ccpiz
interactions are nucleon-level \ccpiz on neutrons, additional smearing is
due to \ccpip charge-exchanges on protons or other inelastic processes
producing a \piz in the final state.  
These smearings are not expected to be large.  The 4-momentum transfer,
$Q^2$, to the hadronic system can be calculated from the reconstructed muon
and neutrino momentum~\cite{nuint09:rhn}.  The calculation of the nucleon
resonance mass is performed from the neutrino and muon momentum also assuming a
stationary neutron (see appendix).  

\section{Event selection}
Isolating observable \ccpiz interactions is challenging as such events are
expected to comprise only 4\% of the data set~\cite{nuance}.  
The sample is dominated by observable \numu-CCQE (44\%), with contributions
from observable \ccpip (19\%), and other CC and NC modes.  Basic sorting is
first performed to separate different classes of 
events based on their PMT hit distributions.  Then the sample is further
refined by cutting on reconstructed quantities to yield an observable \ccpiz
dominated sample.  Each cut is applied and optimized in
succession and will be discuss over the remainder of this section. 

The detector is triggered by a signal from the Booster accelerator
indicating a proton-beam pulse in the Booster neutrino beam line.   
All activity in the detector is recorded for $19.2$~$\mu$s starting
$4.6$~$\mu$s before the $1.6$~$\mu$s neutrino-beam time window.  The detector
activity is grouped into ``subevents:'' clusters in time of PMT hits.   
Groups of 10 or more PMT hits, within a 200~ns window, with
time spacings between the hits of no more than 10~ns with at most two
spacings less than 20~ns, define a subevent~\cite{boone:detnim}.  In an
ideal neutrino event, the first subevent is always caused by the prompt neutrino
interaction; subsequent subevents are due to electrons from stopped-muon
decays.  Neutrino events with one subevent are primarily due to
neutral-current interactions and $\nu_e$-CCQE.  Two-subevent events are from
\numu-CCQE, \ccpiz, and NC\pip.  Three-subevent events are almost completely
\ccpip, with some multi-$\pi$ production. Stopped muon decays produce
electrons with a maximum energy of 53~MeV which never cause more than 200 PMT
hits in the central detector.   

To select a sample of contained events, a requirement is made to reject events
that penetrate the veto.  These events typically cause more than 6 veto PMT hits.
Therefore, a two-subevent sample is defined by requiring more than 
200 tank PMT hits in the first subevent, fewer than 200 tank PMT hits in the second,
and fewer than 6 veto PMT hits for each subevent.  The two-subevent sample
is predicted to be 
71\% \numu-CCQE, 16\% \ccpip, and 6\% \ccpiz.  Observable \ccpip events make it
into the two-subevent sample for several reasons: primarily by \pipcex
and \pipabs in mineral oil, by fast muon decays whose electrons
occur during the prior subevent, and by \mum capture on nucleons affecting 8\%
of \mum in mineral oil.  The two-subevent filter keeps 40\% of CCQE and
\ccpiz interactions while rejecting 80\% of \ccpip interactions.   The
rejection of CCQE and signal is mainly from \mum that exit the tank.
Additionally, events are lost by the 8\% \mum capture rate on carbon.    

\begin{figure}[ht!]
\includegraphics[scale=0.4]{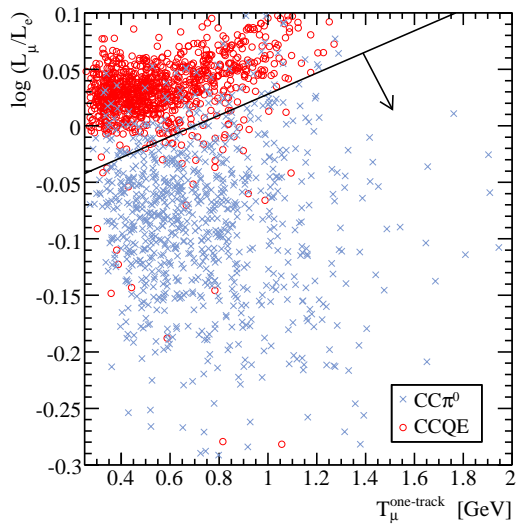}
\caption{(color online) The one-track fit likelihood ratio vs.~the one-track
muon fit kinetic energy.  A separation of CCQE events (red ``o'') relative
to \ccpiz events (grey ``x'') is performed by selecting events below the
black line.  For clarity, the events plotted are prescaled by 1000 and 100
for CCQE and \ccpiz respectively.  The cut is optimized on the full
non-prescaled MC sample.}  
\label{fig:fltr}
\end{figure}
To isolate a purer sample of observable \ccpiz events from the two-subevent
sample, before the observable \ccpiz fit is performed, \numu-CCQE events are
rejected by cutting on the ratio
of the \numu-CCQE fit likelihood to the \nue-CCQE fit likelihood as a
function of \mum-fit kinetic energy (see Fig.~\ref{fig:fltr}).  This cut is
motivated by the fact that \numu-CCQE interactions are dominated by a sharp muon
ring, while \ccpiz interactions (with the addition of two photon rings) will
look ``fuzzier'' and more electron-like to the fitter.   The cut,
which has been optimized to reject CCQE, rejects 96\% of observable
\numu-CCQE while retaining 85\% of observable \ccpiz~\cite{thesis:rhn}.
Additionally, a reconstructed radius cut of $r_\mathrm{rec}<550$ cm is used
to constrain events to within the fiducial volume.  

\begin{figure}[ht!]
\includegraphics[scale=.4]{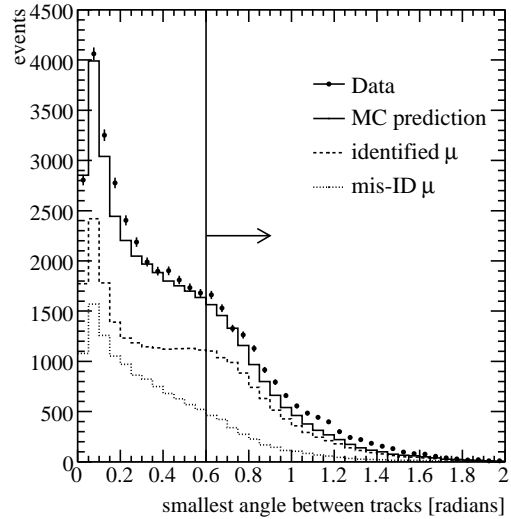}
\caption{The smallest angle between two of the three reconstructed tracks.
Displayed are the data (with statistical errors), total MC (solid),
identified \mum (dashed), and mis-IDed \mum (dotted).  The cut selects
events above 0.6~radians.}
\label{fig:mis}
\end{figure}
To reject misreconstructed signal events, along with certain backgrounds, a
cut is applied to reject events if two of the three \che rings reconstruct
on top of one another.   When this occurs 
the fitter ambiguously divides the total energy between the two tracks.
For cases where there are two or fewer rings, the fitter will place two of
the tracks in the same direction.  For cases where there are
three or more \che rings, the fitter can still get trapped with two tracks on top
of one another.  This can happen either because of asymmetric \piz
decays, a \mum near or below \che threshold, a dominant \mum 
ring, or events that truly have overlapping tracks.  In all of these cases,
the reconstruction becomes poor, especially the identification
of the \mum in the event.  Fig.~\ref{fig:mis} shows the smallest
reconstructed angle between two of the three reconstructed tracks for both
data and MC. The MC events are separated into samples that correctly
identified the \mum and those that did not.  A cut is optimized on signal
events in the MC to reject misidentified \mum, and rejects events with track
separations less than 
0.6~radians~\cite{thesis:rhn}.  This cut reduces the expected \mum
misidentification rate to the 20\% level. Additionally the observable \ccpiz
fraction is increased to 38\% and the observable CCQE fraction is reduced to
13\%.   

\begin{figure}[ht!]
\includegraphics[scale=.4]{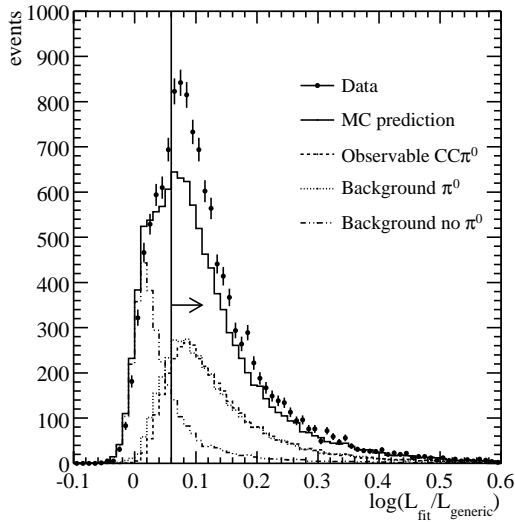}
\caption{The logarithm of the ratio of the observable \ccpiz fit likelihood
to a generic three-track fit.  Displayed are the data (with statistical
errors), total MC (solid), observable \ccpiz (dashed), backgrounds with a \piz
in the final state or produced after the event (dotted), and backgrounds
with no \piz (dot-dashed).  The cut selects events above 0.06.}
\label{fig:like}
\end{figure}
The next series of cuts reject non-\piz backgrounds.  The first
requirement compares the observable \ccpiz fit likelihood vs.~a generic
three-track fit and selects events that are more \ccpiz-like.  
As the observable \ccpiz fit is actually a generic--$\mu\gamma\gamma$ from a
common vertex--fit, this cut selects events that match this criterion.
The second
cut is on the reconstructed $\gamma\gamma$ mass about the expected \piz mass.
This cut demands that the photons are consistent with a
\piz decay. The combination of these cuts define the observable \ccpiz
sample.  Fig.~\ref{fig:like} shows the logarithm of the ratio of the
observable \ccpiz fit over the generic three-track fit likelihoods.  The MC
is separated into three samples: observable \ccpiz, background events with a
\piz in
the final state or created later in the event, and background events with no
\piz.  Both the observable \ccpiz and backgrounds with a \piz are more
$\mu\gamma\gamma$-like than events with no \piz in the event.  Additionally,
as the backgrounds with a \piz either have multiple pions or the \piz was
produced away from the event vertex, the likelihood ratio for these events
tend slightly more toward the generic fit.  Events with no \piz peak sharply
at low ratio values.  The optimization rejects non-\piz backgrounds by selecting
events greater than 0.06 in this ratio~\cite{thesis:rhn}.   
\begin{figure}[ht!]
\includegraphics[scale=0.4]{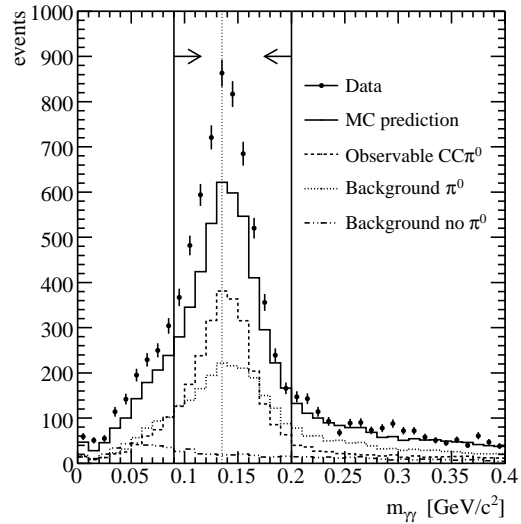}
\caption{The reconstructed $\gamma\gamma$ mass.  Displayed are the data
(with statistical errors), total MC (solid), observable \ccpiz (dashed),
backgrounds with a \piz in the final state or produced after the event
(dotted), and backgrounds with no \piz (dot-dashed).  The vertical dotted
line is the known \piz mass.}
\label{fig:pimass}
\end{figure}

The final cut on the reconstructed $\gamma\gamma$ mass defines the
observable \ccpiz event sample.  Fig.~\ref{fig:pimass} shows the
reconstructed $\gamma\gamma$ mass for both data and MC.  No assumption is
used in the fit that the two photons result from a \piz decay; nevertheless,
both data and MC peak at the known \piz mass.  The predicted background MC with a \piz
in the final state, or a \piz produced after the event, has a broader peak
than the signal MC.  This broadening occurs for the same reasons discussed for
the likelihood ratio; these events either produced a \piz away from the \mum
vertex, or there are multiple pions in the final state.  As one might
expect, background events
with no \piz in the final state show no discernible mass peak and pile up at
low mass with a long misreconstructed tail extending out to high mass.  A
cut is optimized to select events around the known \piz mass [0.09 $<$
$m_{\gamma\gamma}$ $<$ 0.2~GeV] to reject non-\piz backgrounds (low mass
cut) and to increase signal purity (high mass cut)~\cite{thesis:rhn}.   The
addition of these selection cuts increases the observable \ccpiz purity to
57\% with 6\% efficiency.   
\begin{table}
\caption{The expected efficiency and purity of observable \ccpiz events
as a function of applied cut.}
\label{tab:eff}
\begin{tabular}{lrr}
\hline\hline
cut description& \multicolumn{1}{c}{efficiency}    & \multicolumn{1}{c}{purity} \\ \hline
none                                    & 100\%  &  3.6\% \\
Two-subevent and Tank and Veto hits     & 38.2\% &  5.6\% \\
\ccpiz filter and fiducial volume       & 27.9\% & 29.6\% \\
Misreconstruction                       & 10.3\% & 38.1\% \\
Likelihood ratio and $m_{\gamma\gamma}$ &  6.4\% & 57.0\% \\
\hline\hline
\end{tabular}
\end{table}
\begin{table}
\caption{The expected background composition of the \ccpiz candidate
sample by observable mode.  The level of observable \ccpip events are
determined by the method described in Section~\ref{s:ccpip}. 
The symbol $X$ represents all nuclear final
states and photons, and $N\ge 2$ and $M\ge 1$ are the number of pions in the final
state. Other backgrounds include deep inelastic scattering and NC elastic scattering.}
\label{tab:bkgd}
\begin{tabular}{lr@{$\;\to\;$}lr}
\hline\hline
\multicolumn{1}{c}{observable} &\multicolumn{2}{c}{} & \multicolumn{1}{c}{fraction of} \\
\multicolumn{1}{c}{mode} & \multicolumn{2}{c}{description}    & \multicolumn{1}{c}{background}\\ \hline
\ccpip        & $\numu\chtwo$&$\mum\pip X$   & 52.0\% \\
CCQE          & $\numu\chtwo$&$\mum X$       & 15.4\% \\
CCmulti-$\pi$ \quad & $\numu\chtwo$&$\mum(N\pi) X$ & 14.0\% \\
NC-$\pi$      & $\nu_{\phantom{\mu}}\chtwo$&$\nu^{\phantom{-}} (M\pi) X$ &  8.8\% \\
others        & \multicolumn{2}{c}{}         &  9.8\% \\
\hline\hline
\end{tabular}
\end{table}
After all cuts, the observable \ccpiz candidate sample contains 5810 events in data 
for $6.27\times10^{20}$~p.o.t.~while the MC predicts 4160.2
events. Table~\ref{tab:eff} summarizes the effects of 
the cuts on the MC sample,  while Table~\ref{tab:bkgd} summarizes the
background content of the observable \ccpiz candidate sample. 

\section{Analysis}
The extraction of observable \ccpiz cross sections from the event sample
requires a subtraction of background events, corrections for detector effects and
cut efficiencies, a well-understood flux, and an estimation of the number of
interaction targets.  The cross sections are determined by 
\begin{equation}\label{eqn:xsec}
\frac{\partial\sigma}{\partial x}\Big|_i =
\frac{\sum_jU_{ij}(N_j-B_j)}{n\Phi_i\epsilon_i\Delta x_i},
\end{equation}
where $x$ is the variable of interest, $i$ labels a bin of the measurement, 
$\Delta x_i$ is the bin width, $N_j$ is the number of events in data of bin
$j$, $B_j$ is the expected background, $U_{ij}$ is a matrix element that
unfolds out detector effects, $\epsilon_i$ is the bin efficiency, $\Phi_i$
is the predicted neutrino flux, and $n$ is the number of interaction
targets.  For the single-differential cross-section measurements, the flux
factor, $\Phi_i$, is constant and equals the total flux.  For 
the total cross-section measurement as a function of neutrino energy the
flux factor is per energy bin.   
The extracted cross sections require a detailed understanding of the
measurements or predictions of the quantities in Eqn.~\ref{eqn:xsec}, and
their associated systematic uncertainties. 
By construction, the signal cross-section prediction has minimal impact on
the measurements.  Wherever the prediction can affect a measurement, usually
through the systematic error calculations, the dependence is duly
noted.

\subsection{CC$\pi^+$ backgrounds}\label{s:ccpip}
The first stage of the cross-section measurement is to subtract the expected background
contributions from the measured event rate.  This is complicated by
the fact that previously measured modes in MiniBooNE (CCQE~\cite{boone:ccqe:2},
\ncpiz~\cite{boone:ncpi0:2}, and \ccpip~\cite{boone:ccpip}) show substantial
normalization discrepancies with the {\sc Nuance} prediction.  The single largest 
background, observable \ccpip, is well constrained by measurements within the
MiniBooNE data set.  The observable CCQE in the sample is at a small enough
level not warrant further constraint.  Most of the remaining backgrounds are
unmeasured but individually small.  

\begin{figure}[ht!]
\includegraphics[scale=0.4]{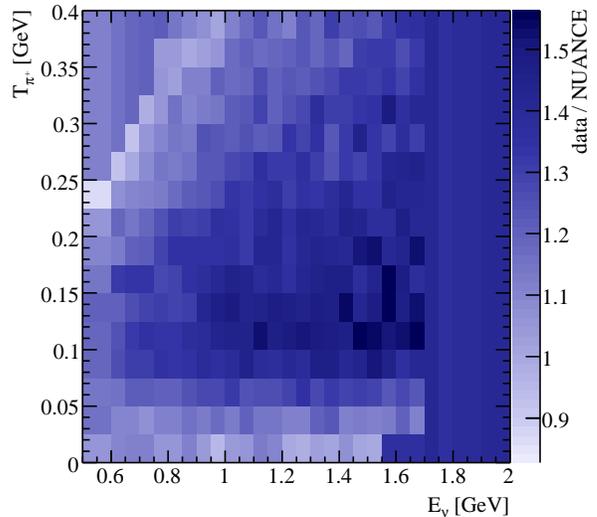}
\caption{(color online) The ratio of the measured observable \ccpip
cross section to the 
{\sc Nuance} prediction.  The ratio is plotted as a function of $T_{\pi^+}$
and $E_\nu$.}
\label{fig:ccpipr}
\end{figure}

The \ccpip backgrounds are important to constrain for two reasons: they
contribute the largest single background, and \pipcex and
\pipabs processes in the mineral oil have large uncertainties.  
In particular, \pipcex in mineral oil is, by definition, not an observable \ccpiz because
the \piz did not originate in the target nucleus.  By tying the
observable \ccpip production to measurements within the MiniBooNE data, 
the uncertainty on this background can also be further reduced.  The total
error is separated into an uncertainty on \ccpip production and an 
uncertainty on \pipcex and \pipabs processes in mineral oil occurring
external to the initial target nucleus. 
Using the high statistics MiniBooNE 3-subevent sample, many measurements of the 
absolute observable \ccpip production cross sections have been performed~\cite{thesis:mike,boone:ccpip}. This 
observable \ccpip sample is predicted to be 90\% pure~\cite{thesis:mike,boone:ccpip} making it the purest 
mode measured in the MiniBooNE data set to date. It is a very useful sample for this 
analysis as the bulk of the \ccpip background events in the \ccpiz sample fall in the 
kinematic region well-measured by the observable \ccpip data. The measurements 
used are from the tables in Ref.~\cite{thesis:mike}. Because the re-interaction cross sections for 
\pipcex and \pipabs are strong functions of \pip energy, the constraint on 
\ccpip events is applied as a function of \pip kinetic energy and neutrino energy. 
Fig.~\ref{fig:ccpipr} shows  
the ratio of the measured observable \ccpip cross section as a function of
\pip and neutrino energies to the {\sc Nuance} predicted cross section.  As
the differential cross section with respect to \pip kinetic energy 
was not reported for bins with small numbers of
events, the reweighting function is patched by the ratio of the total
observable \ccpip cross section as a function of neutrino energy.  This
reweighting factor (as high as 1.6 in some bins) is applied
to every observable \ccpip event in the MC sample.  All figures and numbers
presented in this article have this reweighting applied.  By using the
MiniBooNE data to constrain the \ccpip backgrounds, strict reliance on the
{\sc Nuance} implementation of the Rein-Sehgal model to predict this
important background is avoided.

\subsection{Background subtraction}
The MC predicts a sample that is 57\% pure observable \ccpiz after all
analysis cuts.  A 23\% contribution of observable \ccpip interactions is
set by prior measurements in the MiniBooNE data.  The remaining 20\% of the event
sample is mostly comprised of CCQE and multi-$\pi$ final-state events (see
Table~\ref{tab:bkgd}). These backgrounds, while thought to be produced in
larger quantities than the MC prediction, are set by the MC as
there is no clear method to extract the normalization of many of these modes from
the current data.  If the normalizations are different--but on
the same order as previously measured modes--they would change the final
results by at most a few percent.  
The uncertainties applied to the production of these backgrounds more than
cover the possible normalization differences.  After 
subtracting the backgrounds from the data the sample contains 3725.5 signal
events with $E_\nu\in0.5-2.0$~GeV, which should be compared to 2372.2 predicted events.  
This is a normalization
difference of 1.6, the largest normalization difference that has been
observed in the MiniBooNE data thus far.  
This is attributed to the large effects of pion re-interactions
which can directly influence the measured cross section for observable
\ccpiz production and the fact that the {\sc Nuance} prediction appears low
even when compared to prior measurements on deuterium data~\cite{model:zel}.

\subsection{Unfolding \& flux restriction}
To correct for the effects of detector resolution and reconstruction, 
a method of data unfolding is performed~\cite{cowan}.  The unfolding method
constructs a response matrix from the 
MC that maps reconstructed quantities to their predicted values.  The chosen
method utilizes the Bayesian technique described in
Ref.~\cite{dagostini}.  This method of data unfolding requires a Bayesian
prior for the signal sample, which produces an intrinsic bias.  This bias is
the only way that the signal cross-section model affects the measured cross
sections. In effect, this allows the signal cross-section model to pull the
measured cross sections toward the shape of the distributions produced from
the model while preserving the normalization measured in the data; nevertheless,  
strict dependence on the signal cross-section is avoided.     
For these measurements, the level of uncertainty on the signal model cover
the effect of the bias on the central value.    
For situations where the unfolding is
applied to a distribution that is significantly different than the Bayesian
prior, the granularity of the unfolding matrix needs to be increased to
stabilize the calculation of the systematic uncertainty.  Otherwise the
uncertainties would be larger than expected due to larger intrinsic bias.  
For the total cross section as a function of
neutrino energy, the unfolding acts on the background-subtracted
reconstructed neutrino energy and unfolds back to the neutrino energy prior
to the interaction.  For each of the flux-averaged differential
cross-section measurements, the unfolding acts on the two-dimensional space
of neutrino energy and the reconstructed quantity of a particular
measurement.  For final-state particles, \mum and \piz, the unfolding
corrects to the kinematics after final-state effects, and are the least
model dependent.  For the 4-momentum
transfer, $Q^2$, the unfolding extracts to $Q^2$ calculated from the initial
neutrino and the final-state \mum.  
In both the total cross section and flux-averaged differential
cross section in $Q^2$, the final-state interaction model does bias the
measurements.  Uncertainties in the signal cross-section model, along with
the final-state interaction model, are expected to cover this bias.  
The unfolded two-dimensional distributions are
restricted to a region of unfolded neutrino energy between $0.5-2.0$~GeV.
This effectively restricts the differential cross-section measurements over
the same range of neutrino energy, and flux, as the total cross-section
measurement.  

\subsection{Efficiency correction}
The unfolded distributions are next corrected by a bin-by-bin efficiency.  The
efficiencies are estimated by taking the ratio of MC signal events after
cuts to the predicted distribution of signal events without cuts but restricted
to the fiducial volume.  The efficiency is insensitive to changes in the
underlying MC prediction to within the MC statistical error.  While this
additional statistical error is not large, it is properly accounted for in
the error calculation.  The overall efficiency for selecting observable
\ccpiz interactions is 6.4\%; the bulk of the events ($\sim60\%$) are lost
by demanding a \mum that stops and decays in the MiniBooNE tank.  The cuts
to reduce the backgrounds and preserve well-reconstructed events account for
the remainder.    

\subsection{Neutrino flux}
\begin{figure}[ht!]
  \includegraphics[scale=0.4]{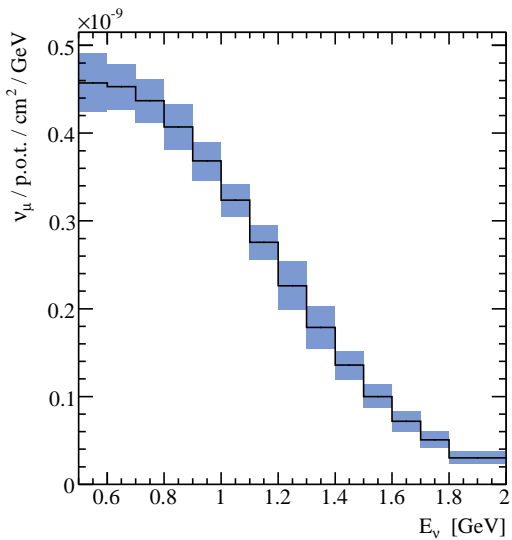}
\caption{(color online) The predicted \numu flux with systematic
uncertainties over the range 0.5--2.0 GeV. A table containing the values of
the flux, along with a full-correlation matrix, is provided in
Table.~\ref{tab:flux} of the appendix.}
\label{fig:flux}
\end{figure}

A major difficulty in extracting absolute cross sections is the need for 
an accurate flux prediction.  The MiniBooNE flux
prediction~\cite{boone:flux} comes strictly from fits to external
data~\cite{harp,e910,aleshin,abbott,allaby,dekkers,eichten,lundy,marmer,vorontsov}
and makes no use of the MiniBooNE neutrino data.
Fig.~\ref{fig:flux} shows the predicted \numu flux with systematic
uncertainties. 
The flux is restricted by the unfolding method to the range 0.5--2.0~GeV,
although contributions from the flux outside of this range affect the systematic
uncertainties of the unfolding method.  The integrated flux over this range
is predicted to be $\Phi_\nu=(3.545\pm0.259)\times10^{-10}$~\numu/p.o.t./cm$^2$.

\subsection{Number of targets}
The Marcol 7 mineral oil is composed of long chains of hydrocarbons.  The
interaction target is one link in a hydrocarbon chain; \chtwo chains with an
additional hydrogen atom at each end.  The molecular weight is the weight of
one unit of the chain averaged over the average chain length.  The density
of the mineral oil is
$\rho_\mathrm{oil}=0.845\pm0.001$~g/cm$^3$~\cite{boone:detnim}.  Thermal
variations over the course of the run 
were less than 1\%.  The fiducial volume is defined to be a sphere 550~cm in
radius.  Therefore, there are $(2.517\pm0.003)\times10^{31}$ interaction targets.  

\section{Systematic uncertainties}
Sources of systematic uncertainties are separated into two types: flux
sources and detector sources.  Flux sources affect the number, type, and
momentum of the neutrino beam at the detector.  Detector sources affect the
interaction of neutrinos in the mineral oil and surrounding materials, the
interactions of the produced particles with the medium, the creation and
propagation of optical photons, and the uncertainties associated with the
detector electronics.  Whenever possible, the input sources to the
systematic uncertainty calculations are fit to both {\it ex}-and {\it
in-situ} data and 
varied within their full error matrices.  These variations are propagated
through the entire analysis chain and determine the systematic error
separately for each quantity of interest.  For sources that affect the
number of interactions (e.g.~flux and cross-section sources), the central
value MC is reweighted by taking the ratio of the value of the underlying
parameter's excursion to that of the central value.  This is done to reduce
statistical variations in the underlying systematics. Sources that change the
properties of an event require separate sets of MC to evaluate the error
matrices.  The unfortunate aspect of this method is it introduces additional
statistical error from the excursion.  If the generated sets are large, then
this additional error is small.  

\subsection{Flux sources}
The main sources of flux uncertainties come from particle production and
propagation in the Booster neutrino beam line.  
The proton-beryllium interactions produce
\pip, \pim, \kp, and \kz particles that decay into \numu neutrinos.  The
dominant source of \numu in the flux range of 0.5--2.0~GeV is \pip decay.
The \pip production uncertainties are determined by propagating the HARP
measurement error matrix by spline interpolation and reweighting into an
error matrix describing the contribution of \pip production to the
MiniBooNE \numu flux uncertainty~\cite{boone:flux}. 
Over the flux range of this analysis, the total uncertainty on \pip
production introduces a 6.6\% uncertainty on the \numu-flux.  These
uncertainties would be larger if the entire neutrino flux were considered in
these measurements; however, as the largest flux uncertainties are in
regions of phase space that cannot produce a \ccpiz interaction, it was
prudent to restrict the flux range. Horn related uncertainties stem
from horn-current variations 
and skin-depth effects (which mainly affect the high-energy neutrino flux), 
along with other beam related effects (e.g. secondary interactions), 
and provide an additional 3.8\% uncertainty. No other source contributes
more than 0.2\% (\kp production) to the uncertainty.  
The total uncertainty on the flux over the range 0.5--2.0~GeV, from
all flux-related sources, is 7.3\%.  
As the flux prediction affects the measured
cross sections through the flux weighting, background subtraction, and
unfolding, the resulting uncertainties on the measured total cross section
are 7.5\% and 7.3\% for horn variations and \pip production respectively.   

\subsection{Detector sources}
Uncertainties associated with the detector result from: neutrino-interaction
cross sections, charged-particle interactions 
in the mineral oil, the creation and interaction of photons in the mineral
oil, and the detector read-out electronics. The neutrino-interaction
cross sections are varied in {\sc Nuance} within their error matrices.
These variations mainly affect the background predictions; however, the
expected signal variations (i.e.~flux and signal cross-section variations)
do affect the unfolding.  The variations of the 
background cross sections cover the differences seen in the MiniBooNE data; the
variations of the signal cover the bias introduced through the chosen
unfolding method.  The uncertainty on the observable \ccpip cross section is
constrained by measurements within the data~\cite{thesis:mike,boone:ccpip} 
assuming no bin-to-bin correlations.  The total cross-section uncertainty on
the observable \ccpiz cross-section measurement is 5.8\%. 

The creation and propagation of optical photons in the mineral oil is
referred to as the optical model.  Several {\em ex-situ} measurements were
performed on the mineral oil to accurately describe elastic Rayleigh and 
inelastic Raman scattering, along with the fluorescence
components~\cite{boone:detnim,brown:2004uy}.  Additionally, reflections and PMT
efficiencies are included in the model, which is defined by a total of 35
correlated parameters.  
These parameters are varied, in a correlated
manner, over a set of data-sized MC samples.  The uncertainty calculated
from these MC samples contains an additional amount of statistical error; however,
in this analysis, the bulk of the additional statistical error is smoothed
out by forcing each MC to have the same underlying true distributions.  The
optical model uncertainty on the total cross-section measurement 
is 2.8\%.  

Variations in the detector electronics are estimated as PMT
effects.  The first measures the PMT response by adjusting the discriminator
threshold from 0.1~photoelectrons (PE) to 0.2~PE.  The second measures the
correlation between the charge and time of the PMT hits.  These
uncertainties contribute 5.7\% and 1.1\% to the total cross section respectively.  

The dominant uncertainty on the cross-section measurements comes from the
uncertainty of \pipcex and \pipabs in the mineral oil occurring external to the
initial target nucleus where the effects of
\pipcex and \pipabs internal the target nucleus are included in these 
measurements.  The uncertainty on
the \pipcex cross section on mineral oil is 50\% and for \pipabs it is 35\%.  The
uncertainties come from external data~\cite{exp:ashery}.   These uncertainties
affect this measurement to a large degree because of the much larger observable
\ccpip interaction rate (by a factor of 5.2).  While this uncertainty is small
in the observable \ccpip cross-section measurements, many \ccpip that
undergo either \pipcex or \pipabs in the mineral oil end up in the \ccpiz
candidate sample. The uncertainty on the \ccpiz cross sections of \pipcex
and \pipabs from observable \ccpip interactions is 12.9\%.  The uncertainty
applied to CC(NC)multi-$\pi$, \ncpip, and other backgrounds is included in
the background cross-section uncertainties.   

\subsection{Discussion}
The total systematic uncertainty, from all sources, on the observable \ccpiz
total cross-section measurement is 18.7\%.  The total uncertainty is found
by summing all of the individual error matrices.  The largest uncertainty,
\pipcex and \pipabs in the mineral oil, is 12.9\%; the flux uncertainties are 10.5\%; the
remaining detector and neutrino cross-section uncertainties are 8.6\%.  The
total statistical uncertainty is 3.3\%.  Table~\ref{tab:xsecsum} summarizes
the effects of all sources of systematic uncertainty.  Clearly, the limiting
factor on the measurement is the understanding of \pipcex and \pipabs in
mineral oil external to the target nucleus.
The two simplest ways to reduce this uncertainty in future experiments are
to improve the understanding of pion scattering in a medium, or to use a
fine-grained detector that can observe the \pip before the charge-exchange or
absorption.  Beyond that, gains can always be made from an improved
understanding of the incoming neutrino flux.   

\begin{table}[ht!]
\caption{Summary of systematic uncertainties.}  
\label{tab:xsecsum}
\begin{tabular}{lr@{.}l}\hline\hline
source & \multicolumn{2}{c}{uncertainty} \\ \hline
\pipcex\& \pipabs in mineral oil & \pz\pz 12&9\% \\  
\numu flux                       & \pz\pz 10&5\% \\ 
$\nu$ cross section              & \pz\pz  5&8\% \\ 
detector electronics             & \pz\pz  5&8\% \\ 
optical model                    & \pz\pz  2&8\% \\  \hline
total                            & \pz\pz 18&7\% \\  \hline\hline
\end{tabular}
\end{table}

\section{Results}
This report presents measurements of the observable \ccpiz cross section as
a function of neutrino energy,  
and flux-averaged differential cross sections in $Q^2$, $E_\mu$,
$\cos\theta_\mu$, $|\mathbf{p}_{\piz}|$, and $\cos\theta_{\piz}$.  These
measurements provide the most complete information about this interaction on
a nuclear target (\chtwo) at these energies (0.5--2.0~GeV) to date.  
Great care has been taken to measure cross sections with minimal
neutrino-interaction model dependence.  First, the definition of
an observable \ccpiz interaction limits the dependence of these
measurements on the internal \pipcex and \piabs models. Second, most of the
measurements are presented in terms of the observed final-state particle
kinematics further reducing the dependence of the measurements on the FSI
model. Any exceptions to these are noted with the measurements.

\begin{figure}[h!]
\includegraphics[scale=0.4]{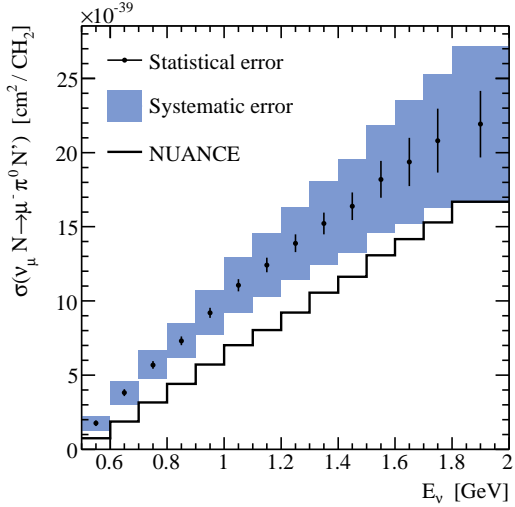}
\caption{(color online) The total observable \ccpiz cross section as a
function of neutrino energy.  The uncertainty is dominated by \pip
charge-exchange and absorption in the mineral oil external the target
nucleus.  The total systematic uncertainty on the cross section is 18.7\%.
The central-value measurement, uncertainties, and correlation matrix are
tabulated in Table~\ref{tab:xsec} of the appendix.}
\label{fig:xsec}
\end{figure}

The first result, a measurement of the total observable \ccpiz cross section is
shown in Fig.~\ref{fig:xsec}.  This measurement is performed by integrating
Eqn.~\ref{eqn:xsec} over neutrino energy.  This result does contain some
dependence on the initial neutrino-interaction model as the unfolding
extracts back to the initial neutrino energy; however, this dependence is
not expected to be large.  The total measured systematic uncertainty is
18.7\%, and is slightly higher than the uncertainties presented for the
cross sections without this dependence. 
The total cross section is higher at all energies than is expected 
from the combination of the initial interaction~\cite{model:RS} and FSI as
implemented in {\sc Nuance}.  An enhancement is also
observed in other recent charged-current cross-section
measurements~\cite{boone:ccqe:2,boone:ccpip}; however, the enhancement is a
factor of $1.56\pm0.26$ larger than the prediction here.   

\begin{figure}[h!]
\includegraphics[scale=0.4]{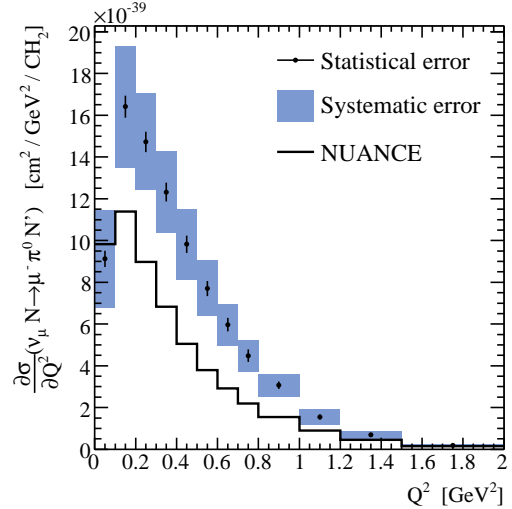}
\caption{(color online) The flux-averaged ($0.5<E_\nu<2.0$~GeV)
differential cross section in $Q^2$ with  total systematic uncertainty of
16.6\%.  The central-value measurement, uncertainties, and correlation matrix
are tabulated in Table~\ref{tab:dxsecdq2} of the appendix.} 
\label{fig:dxsecdq2}
\end{figure}

The differential cross sections provide additional insight into the effect of
final-state interactions.  By necessity, these cross sections are presented
as flux-averaged results due to the low statistics of these measurements.  Except
for the measurement in $Q^2$, these measurements are mostly independent of the
underlying neutrino-interaction model (though there is a slight influence
from the unfolding). The flux-averaged cross section, differential in $Q^2$
(Fig.~\ref{fig:dxsecdq2}), is dependent on the initial neutrino-interaction
model because it requires knowledge of the initial neutrino kinematics.  The
measured $Q^2$ is unfolded to the $Q^2$ calculated from the initial neutrino and
final-state muon kinematics.  This measurement 
shows an overall enhancement along with a low-$Q^2$ suppression (relative to
the normalization difference) with a total
systematic uncertainty of 16.6\%.  A similar disagreement is also observed
in the \ccpip cross-section measurement~\cite{boone:ccpip}.  

\begin{figure}[h!]
\includegraphics[scale=0.4]{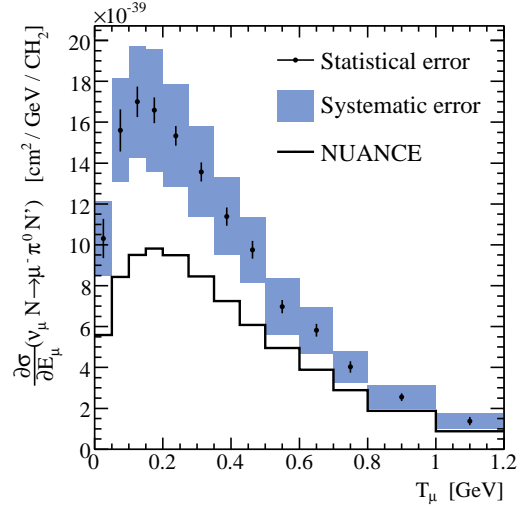}
\caption{(color online) The flux-averaged ($0.5<E_\nu<2.0$~GeV) differential cross section in
$E_\mu$ with total systematic uncertainty of 15.8\%. The central-value
measurement, uncertainties, and correlation matrix are tabulated in
Table~\ref{tab:dxsecdemu} of the appendix.} 
\label{fig:dxsecdemu}
\end{figure}

\begin{figure}[h!]
\includegraphics[scale=0.4]{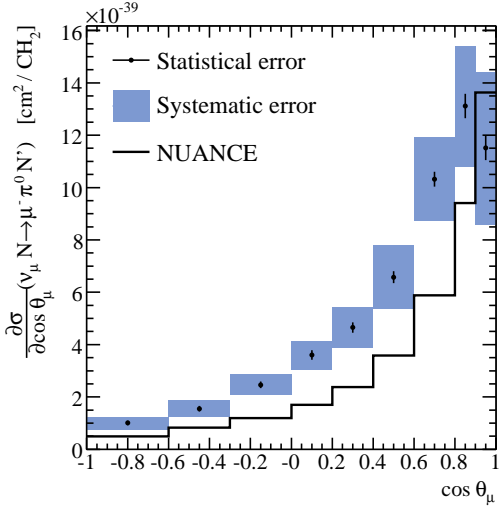}
\caption{(color online) The flux-averaged ($0.5<E_\nu<2.0$~GeV) differential cross section in
$\cos\theta_\mu$ with total systematic uncertainty of 17.4\%.  The
central-value measurement, uncertainties, and correlation matrix are
tabulated in Table~\ref{tab:dxsecdcosmu} of the appendix.}
\label{fig:dxsecdcosmu} 
\end{figure}

The kinematics of the \mum are fully specified by its kinetic energy and
angle with respect to the incident neutrino beam as the beam is unpolarized.  
Fig.~\ref{fig:dxsecdemu}
shows the flux-averaged differential cross section in \mum kinetic energy.  Like the
total cross section, this shows primarily an effect of an overall enhancement of the
cross section as the \mum is not expected to be subject to final-state
effects.  The total systematic uncertainty for this measurement is 15.8\%.  The 
flux-averaged differential cross section in \mum-\numu angle
(Fig.~\ref{fig:dxsecdcosmu}),  
shows a suppression of the cross section at forward angles,
characteristic of the low-$Q^2$ suppression.  
As the shapes of the
data and the Rein-Sehgal model as implemented in {\sc Nuance} are
fundamentally different in this variable, the unfolding procedure required
a reduction of the number of bins in order to be stable. 
The total systematic uncertainty is 17.4\%.  

\begin{figure}[h!]
\includegraphics[scale=0.4]{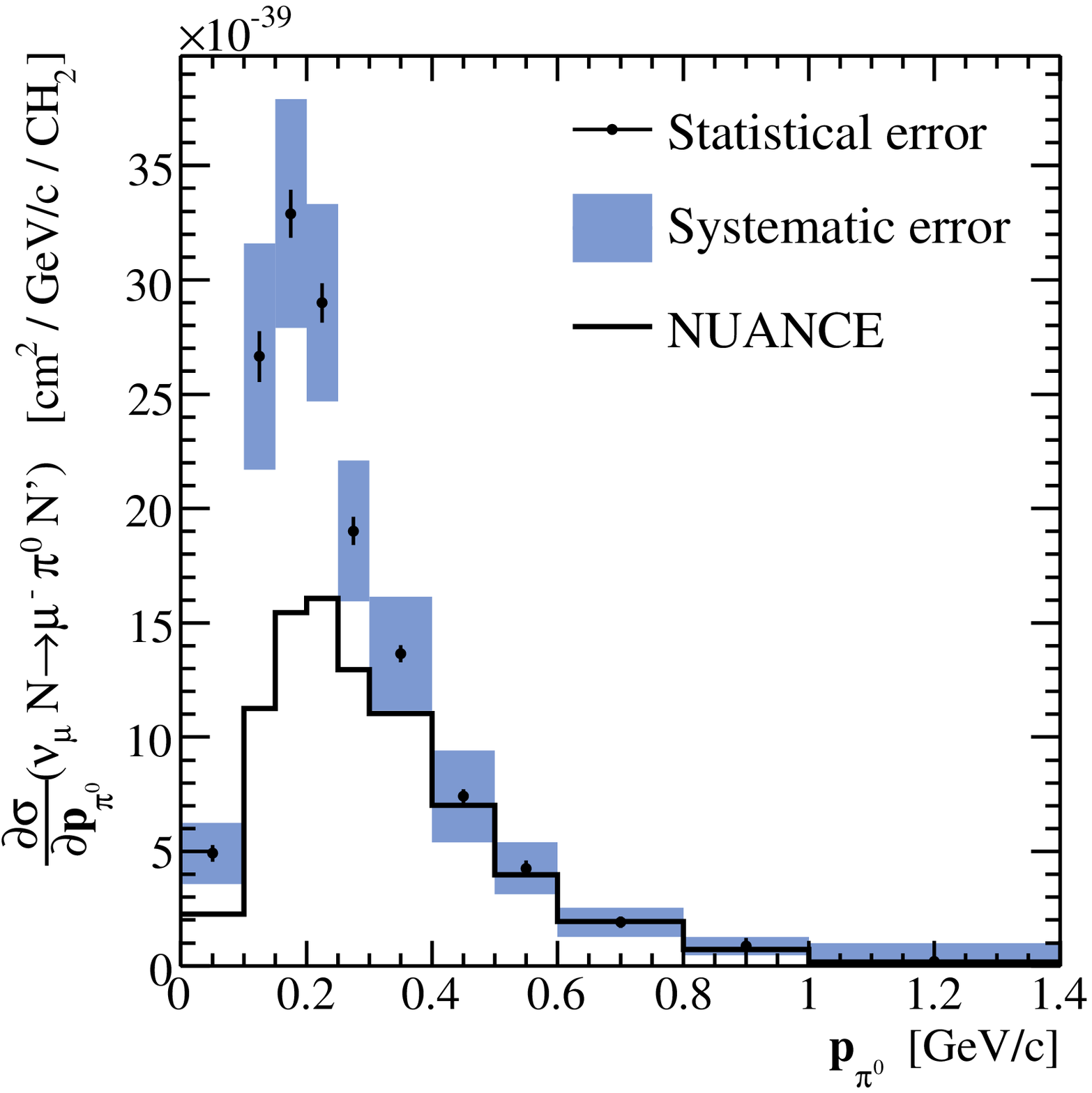}
\caption{(color online) The flux-averaged ($0.5<E_\nu<2.0$~GeV) differential cross section in 
$|\mathbf{p}_{\piz}|$ with total systematic uncertainty of 15.9\%. The
central-value measurement, uncertainties, and correlation matrix are
tabulated in Table~\ref{tab:dxsecdppi} of the appendix.}
\label{fig:dxsecdppi} 
\end{figure}

\begin{figure}[h!]
\includegraphics[scale=0.4]{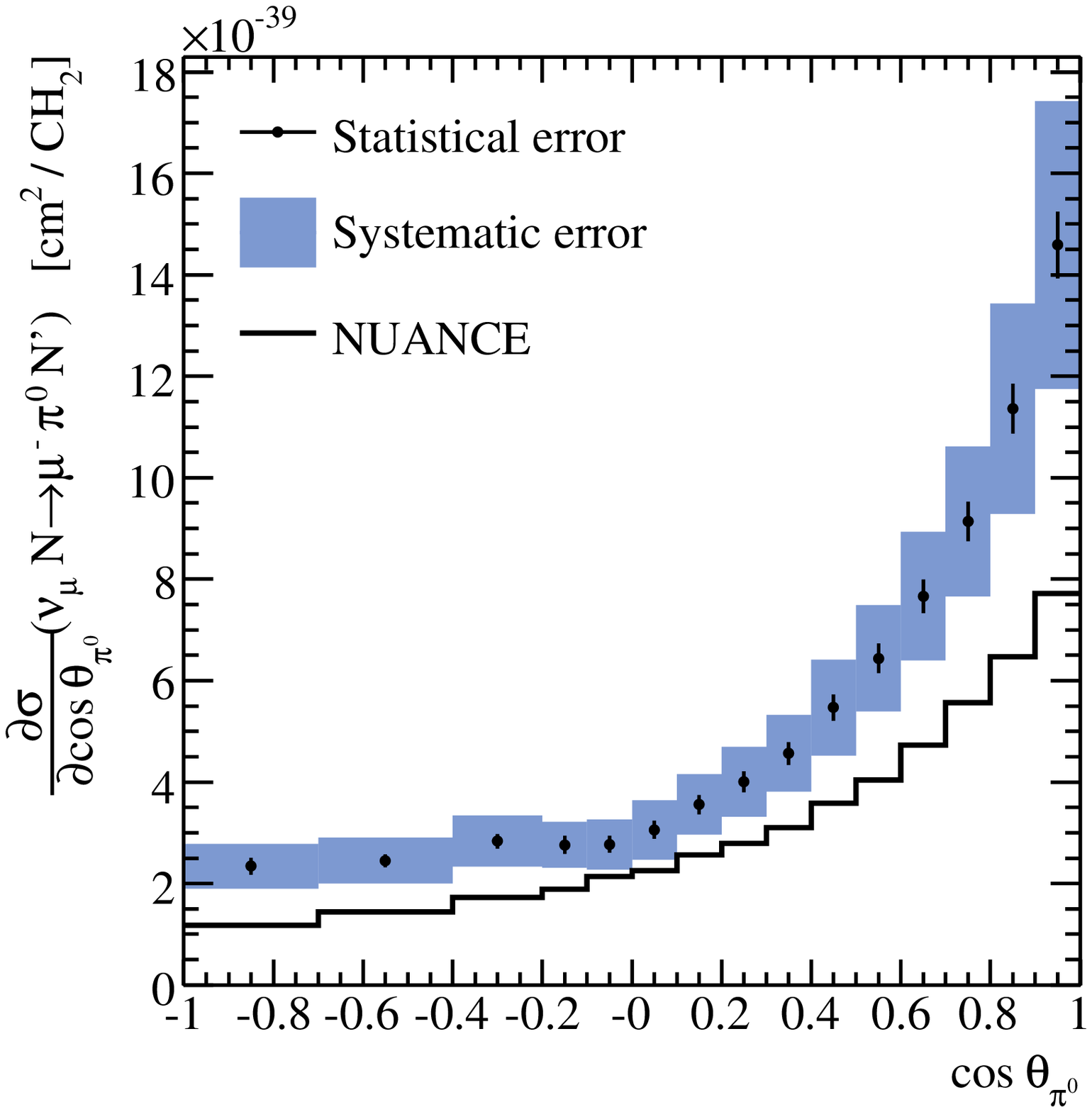}
\caption{(color online) The flux-averaged ($0.5<E_\nu<2.0$~GeV) differential cross section in
$\cos\theta_{\piz}$ with total systematic uncertainty of 16.3\%. The
central-value measurement, uncertainties, and correlation matrix are
tabulated in Table~\ref{tab:dxsecdcospi} of the appendix.} 
\label{fig:dxsecdcospi}
\end{figure}

The \piz kinematics yield insight into the final-state interaction effects
and are also fully specified by two measurements: the pion momentum and
angle with respect to the neutrino beam direction.  
Fig.~\ref{fig:dxsecdppi}  
shows the flux-averaged differential cross section in
$|\mathbf{p}_{\piz}|$.  The cross section is enhanced at low momentum and in
the peak, but agrees with the prediction at higher momentum.  A similar
disagreement is also observed in the \ncpiz cross-section
measurements~\cite{boone:ncpi0:2}. Interactions
of both the nucleon resonance and pions with the nuclear medium can cause
the ejected \piz to have lower momentum.  The total systematic uncertainty
is 15.9\%.  Fig.~\ref{fig:dxsecdcospi} shows the flux-averaged 
differential cross section in \piz-\numu angle.  The cross section is more
forward than the prediction.  The total systematic error is 16.3\%.   

Each of the cross-section measurements also provide a measurement of the
flux-averaged total cross section.  From these total cross sections, all of the
observable \ccpiz cross-section measurements can be compared.
Table~\ref{tab:summary} 
\begin{table}[ht!]
\caption{Summary of the flux-averaged total cross sections calculated from
each cross-section measurement.  The average cross section is calculated
assuming 100\% correlated systematics. The flux-averaged neutrino energy is
$\braket{E_\nu}_\Phi=0.965$~GeV.}  
\label{tab:summary}
\begin{tabular}{lr@{$\pm$}l}\hline\hline
measurement & \multicolumn{2}{c}{$\braket{\sigma}_\Phi$ [$\times10^{-39}\;$cm$^2$]} \\ \hline
$\sigma(E_\nu)$                                   & \pz\pz9.05&1.44 \\  
$\partial\sigma/\partial Q^2$                     & \pz\pz9.28&1.55 \\ 
$\partial\sigma/\partial E_\mu$                   & \pz\pz9.20&1.47 \\ 
$\partial\sigma/\partial \cos\theta_\mu$          & \pz\pz9.10&1.50 \\  
$\partial\sigma/\partial|\mathbf{p}_{\pi^0}|$     & \pz\pz9.03&1.54 \\ 
$\partial\sigma/\partial \cos\theta_{\pi^0}$      & \pz\pz9.54&1.55 \\ \hline
\multicolumn{1}{c}{$\braket{\bar{\sigma}}_\Phi$} & \pz\pz9.20&1.51 \\ \hline\hline
\end{tabular}
\end{table}
shows the flux-averaged total cross sections calculated from each
measurement.  The measurements all agree within 6\%, well within the
uncertainty.  While all measurements use the same data, small differences
can result from biases due to the efficiencies and
unfoldings.  The results are combined in a simple average, assuming 100\%
correlated uncertainties, to yield 
$\braket{\bar{\sigma}}_\Phi=(9.20\pm0.3_{stat.}\pm1.51_{syst.})\times10^{-39}$~cm$^2$/\chtwo
at flux-averaged neutrino energy of $\braket{E_\nu}_\Phi=0.965$~GeV.  The
averaged flux-averaged total cross-section measurement is found to be a
factor of $1.58\pm0.05_{stat.}\pm0.26_{syst.}$ higher than the {\sc Nuance}
prediction.  

\section{Conclusion}
The measurements presented here provide the most complete understanding of \ccpiz
interactions at energies below 2 GeV to date.  
They are the first on a nuclear target (\chtwo), at these energies, and
provide differential cross-section  
measurements in terms of the final state, non-nuclear, particle kinematics.  
The development of a novel 3-\che ring fitter has facilitated the
reconstruction of both the \piz and \mum in a \ccpiz interaction.  This
reconstruction allows for the measurement of the full kinematics of the
event providing for the measurement of six cross sections: the total
cross section as a function of 
neutrino energy, and flux-averaged differential cross sections in $Q^2$,
$E_\mu$, $\cos\theta_\mu$, $|\mathbf{p}_{\piz}|$, and $\cos\theta_{\piz}$.
These cross sections show an enhancement over the initial interaction
model~\cite{model:RS} and FSI effects as implemented in {\sc Nuance}.  
The flux-averaged total cross section is measured to be
$\braket{\bar{\sigma}}_\Phi=(9.20\pm0.3_{stat.}\pm1.51_{syst.})\times10^{-39}$~cm$^2$/\chtwo
at mean neutrino energy of $\braket{E_\nu}_\Phi=0.965$~GeV.  
These measurements should prove useful for understanding
incoherent pion production on nuclear targets.  

\begin{acknowledgements} 
The authors would like to acknowledge the support of Fermilab, the
Department of Energy, and the National Science Foundation in the
construction, operation, and data analysis of the Mini Booster Neutrino
Experiment.  
\end{acknowledgements} 


\appendix*

\section{Hadronic invariant mass}
\begin{figure}[ht!]
\includegraphics[scale=0.4]{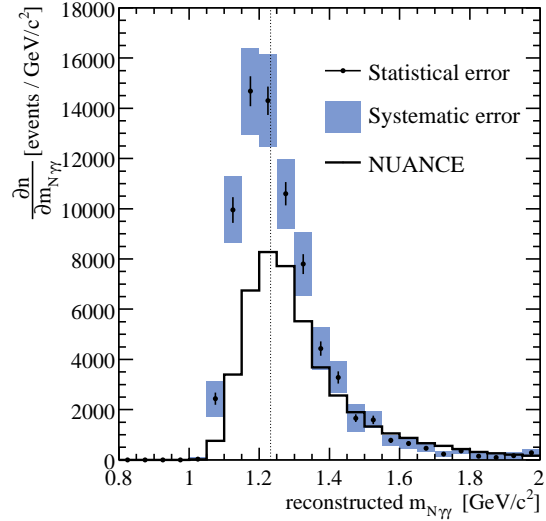}
\caption{(color online) Invariant mass of the hadronic system for both data
(points with error bars) and the MC prediction (solid line) for the signal
mode.  The dotted line indicates the location of the $\Delta(1232)$ resonance.}
\label{fig:delta}
\end{figure}

The background-subtracted reconstructed nucleon resonance mass is calculable
from the reconstructed neutrino and muon 4-momenta.   
Fig.~\ref{fig:delta} shows the
background-subtracted reconstructed   
invariant mass for data and the MC expectation.  The data has not been
corrected for cut efficiencies.   The fact that the
data peaks somewhat below the $\Delta(1232)$ resonance while {\sc Nuance} peaks at
the resonance implies the model is not properly taking into account final state
interaction effects; however, it is observed that the MiniBooNE data is
almost completed dominated by the $\Delta(1232)$ resonance.  This shift can
also be interpreted as an effective change in 
the recoil mass, $W$,
of the hadronic system. Additionally, it has been verified that CCQE
interactions, which do not involve a resonance, peak at threshold
(not displayed).

\section{Tables}
The tables presented in this appendix are provided to quantify the flux,
Fig.~\ref{fig:flux}, and the cross-section measurements,
Figs.~\ref{fig:xsec}, \ref{fig:dxsecdq2}, \ref{fig:dxsecdemu},
\ref{fig:dxsecdcosmu},  \ref{fig:dxsecdppi}, and \ref{fig:dxsecdcospi}.

\begin{turnpage}
\begin{table}[h!]
\caption{The predicted \numu flux.  Tabulated are the central value, total
systematic uncertainty, and the correlation matrix.
The bin boundaries are tabulated as the low edge of the bin with the last 
bin giving both the low and high edge.}
\label{tab:flux}
\begin{tabular}{c|r@{.}lr@{.}lr@{.}lr@{.}lr@{.}lr@{.}lr@{.}lr@{.}lr@{.}lr@{.}lr@{.}lr@{.}lr@{.}lr@{.}l}
Bin edge [GeV] & 0&50 & 0&60 & 0&70 & 0&80 & 0&90 & 1&00 & 1&10 & 1&20 & 1&30 & 1&40 & 1&50 & 1&60 & 1&70 & \multicolumn{2}{c}{1.80-2.00} \\ \hline\hline
CV [$\times10^{-10}\;\nu$/p.o.t./cm$^2$/GeV]& 4&57 & 4&53 & 4&37 & 4&07 & 3&68 & 3&24 & 2&76 & 2&26 & 1&79 & 1&36 & 1&00 & 0&72 & 0&51 & 0&30 \\
Total Syst. & $\pm$0&33 & $\pm$0&26 & $\pm$0&25 & $\pm$0&26 & $\pm$0&22 &
$\pm$0&18 & $\pm$0&20 & $\pm$0&28 & $\pm$0&24 & $\pm$0&17 & $\pm$0&13 &
$\pm$0&12 & $\pm$0&10 & $\pm$0&07 \\ \hline
0.50      &  1&00 &  0&91 &  0&91 &  0&79 &  0&77 &  0&45 & -0&06 & -0&35 & -0&33 & -0&12 & 0&12 & 0&19 & 0&25 & 0&28  \\
0.60      &  0&91 &  1&00 &  0&92 &  0&80 &  0&77 &  0&54 &  0&08 & -0&16 & -0&16 & -0&01 & 0&13 & 0&15 & 0&17 & 0&18  \\
0.70      &  0&91 &  0&92 &  1&00 &  0&87 &  0&86 &  0&61 &  0&13 & -0&19 & -0&18 &  0&04 & 0&25 & 0&28 & 0&31 & 0&33  \\
0.80      &  0&79 &  0&80 &  0&87 &  1&00 &  0&92 &  0&69 &  0&15 & -0&17 & -0&17 &  0&09 & 0&35 & 0&40 & 0&41 & 0&44  \\
0.90      &  0&77 &  0&77 &  0&86 &  0&92 &  1&00 &  0&83 &  0&34 & -0&02 & -0&01 &  0&23 & 0&44 & 0&47 & 0&49 & 0&50  \\
1.00      &  0&45 &  0&54 &  0&61 &  0&69 &  0&83 &  1&00 &  0&75 &  0&46 &  0&44 &  0&60 & 0&65 & 0&62 & 0&57 & 0&55  \\
1.10      & -0&06 &  0&08 &  0&13 &  0&15 &  0&34 &  0&75 &  1&00 &  0&89 &  0&87 &  0&87 & 0&70 & 0&58 & 0&47 & 0&38  \\
1.20      & -0&35 & -0&16 & -0&19 & -0&17 & -0&02 &  0&46 &  0&89 &  1&00 &  0&98 &  0&84 & 0&54 & 0&40 & 0&28 & 0&19  \\
1.30      & -0&33 & -0&16 & -0&18 & -0&17 & -0&01 &  0&44 &  0&87 &  0&98 &  1&00 &  0&91 & 0&63 & 0&49 & 0&38 & 0&26  \\
1.40      & -0&12 & -0&01 &  0&04 &  0&09 &  0&23 &  0&60 &  0&87 &  0&84 &  0&91 &  1&00 & 0&89 & 0&78 & 0&66 & 0&53  \\
1.50      &  0&12 &  0&13 &  0&25 &  0&35 &  0&44 &  0&65 &  0&70 &  0&54 &  0&63 &  0&89 & 1&00 & 0&97 & 0&89 & 0&79  \\
1.60      &  0&19 &  0&15 &  0&28 &  0&40 &  0&47 &  0&62 &  0&58 &  0&40 &  0&49 &  0&78 & 0&97 & 1&00 & 0&97 & 0&91  \\
1.70      &  0&25 &  0&17 &  0&31 &  0&41 &  0&49 &  0&57 &  0&47 &  0&28 &  0&38 &  0&66 & 0&89 & 0&97 & 1&00 & 0&97  \\
1.80-2.00 &  0&28 &  0&18 &  0&33 &  0&44 &  0&50 &  0&55 &  0&38 &  0&19 &  0&26 &  0&53 & 0&79 & 0&91 & 0&97 & 1&00
\end{tabular}
\end{table}
\end{turnpage}

\begin{turnpage}
\begin{table}[h!]
\caption{The total observable \ccpiz cross section as a function of
neutrino energy (Fig.~\ref{fig:xsec}).  Tabulated are the central value
(CV), the total systematic error, the statistical error, and the correlation
matrix for the systematic error.  The
correlations for the statistical errors are small and not tabulated.  The CV,
total systematic error, and statistical errors are multiplied by $10^{39}$.
The bin boundaries are tabulated as the low edge of the bin with the last
bin giving both the low and high edge.}
\label{tab:xsec}
\begin{tabular}{c|r@{.}lr@{.}lr@{.}lr@{.}lr@{.}lr@{.}lr@{.}lr@{.}lr@{.}lr@{.}lr@{.}lr@{.}lr@{.}lr@{.}l}
Bin edge [GeV] & 0&50 & 0&60 & 0&70 & 0&80 & 0&90 & 1&00 & 1&10 & 1&20 & 1&30 & 1&40 & 1&50 & 1&60 & 1&70 & \multicolumn{2}{c}{1.80-2.00} \\ \hline\hline
CV [$\times10^{-39}\;$cm$^2$]& 1&76 & 3&83 & 5&68 & 7&31 & 9&20 & 11&06 & 12&42 & 13&89 & 15&23 & 16&38 & 18&20 & 19&37 & 20&80 & 21&92 \\
Stat. & $\pm$0&18 & $\pm$0&23 & $\pm$0&26 & $\pm$0&29 & $\pm$0&34 & $\pm$0&41 & $\pm$0&49 & $\pm$0&60 & $\pm$0&74 & $\pm$0&92 & $\pm$1&24 & $\pm$1&62 & $\pm$2&16 & $\pm$2&24 \\
Total Syst. & $\pm$0&49 & $\pm$0&78 & $\pm$0&97 & $\pm$1&17 & $\pm$1&50 & $\pm$1&85 & $\pm$2&16 & $\pm$2&46 & $\pm$2&82 & $\pm$3&17 & $\pm$3&61 & $\pm$4&15 & $\pm$4&49 & $\pm$5&26 \\ \hline
0.50 & 1&00 & 0&93 & 0&80 & 0&74 & 0&71 & 0&77 & 0&74 & 0&68 & 0&63 & 0&59 & 0&50 & 0&50 & 0&49 & 0&54  \\
0.60 & 0&93 & 1&00 & 0&96 & 0&90 & 0&88 & 0&86 & 0&81 & 0&79 & 0&72 & 0&68 & 0&62 & 0&56 & 0&59 & 0&59  \\
0.70 & 0&80 & 0&96 & 1&00 & 0&98 & 0&96 & 0&92 & 0&85 & 0&86 & 0&79 & 0&75 & 0&73 & 0&64 & 0&68 & 0&65  \\
0.80 & 0&74 & 0&90 & 0&98 & 1&00 & 0&99 & 0&96 & 0&91 & 0&91 & 0&87 & 0&84 & 0&81 & 0&75 & 0&77 & 0&74  \\
0.90 & 0&71 & 0&88 & 0&96 & 0&99 & 1&00 & 0&97 & 0&93 & 0&94 & 0&89 & 0&86 & 0&85 & 0&77 & 0&80 & 0&75  \\
1.00 & 0&77 & 0&86 & 0&92 & 0&96 & 0&97 & 1&00 & 0&99 & 0&98 & 0&95 & 0&92 & 0&89 & 0&85 & 0&86 & 0&83  \\
1.10 & 0&74 & 0&81 & 0&85 & 0&91 & 0&93 & 0&99 & 1&00 & 0&99 & 0&97 & 0&96 & 0&91 & 0&91 & 0&90 & 0&89  \\
1.20 & 0&68 & 0&79 & 0&86 & 0&91 & 0&94 & 0&98 & 0&99 & 1&00 & 0&99 & 0&97 & 0&95 & 0&91 & 0&92 & 0&89  \\
1.30 & 0&63 & 0&72 & 0&79 & 0&87 & 0&89 & 0&95 & 0&97 & 0&99 & 1&00 & 0&99 & 0&97 & 0&96 & 0&95 & 0&94  \\
1.40 & 0&59 & 0&68 & 0&75 & 0&84 & 0&86 & 0&92 & 0&96 & 0&97 & 0&99 & 1&00 & 0&98 & 0&98 & 0&97 & 0&96  \\
1.50 & 0&50 & 0&62 & 0&73 & 0&81 & 0&85 & 0&89 & 0&91 & 0&95 & 0&97 & 0&98 & 1&00 & 0&97 & 0&97 & 0&94  \\
1.60 & 0&50 & 0&56 & 0&64 & 0&75 & 0&77 & 0&85 & 0&91 & 0&91 & 0&96 & 0&98 & 0&97 & 1&00 & 0&97 & 0&98  \\
1.70 & 0&49 & 0&59 & 0&68 & 0&77 & 0&80 & 0&86 & 0&90 & 0&92 & 0&95 & 0&97 & 0&97 & 0&97 & 1&00 & 0&97  \\
1.80-2.00 & 0&54 & 0&59 & 0&65 & 0&74 & 0&75 & 0&83 & 0&89 & 0&89 & 0&94 & 0&96 & 0&94 & 0&98 & 0&97 & 1&00
\end{tabular}
\end{table}
\end{turnpage}

\begin{turnpage}
\begin{table}[h!]
\caption{The flux-averaged observable \ccpiz differential cross section
 in $Q^2$ over the flux range $E_\nu\in(0.5-2.0\;$GeV$)$
(Fig.~\ref{fig:dxsecdq2}).  Tabulated are the central value 
(CV), the total systematic error, the statistical error, and the correlation
matrix for the systematic error.  The
correlations for the statistical errors are small and not tabulated.  The CV,
total systematic error, and statistical errors are multiplied by
$10^{39}$. The bin boundaries are tabulated as the low edge of the bin with
the last bin giving both the low and high edge.}
\label{tab:dxsecdq2}
\begin{tabular}{c|r@{.}lr@{.}lr@{.}lr@{.}lr@{.}lr@{.}lr@{.}lr@{.}lr@{.}lr@{.}lr@{.}lr@{.}l}
Bin edge [GeV$^2$] & 0&00 & 0&10 & 0&20 & 0&30 & 0&40 & 0&50 & 0&60 & 0&70 & 0&80 & 1&00 & 1&20 & \multicolumn{2}{c}{1.50-2.00} \\ \hline\hline
CV [$\times10^{-39}\;$cm$^2$/GeV$^2$] & 9&12 & 16&41 & 14&72 & 12&32 & 9&82 & 7&70 & 5&97 & 4&48 & 3&07 & 1&55 & 0&70 & 0&18 \\
Stat. & $\pm$0&39 & $\pm$0&52 & $\pm$0&49 & $\pm$0&45 & $\pm$0&41 & $\pm$0&36 & $\pm$0&33 & $\pm$0&30 & $\pm$0&17 & $\pm$0&12 & $\pm$0&07 & $\pm$0&02 \\ 
Total Syst. & $\pm$2&33 & $\pm$2&90 & $\pm$2&31 & $\pm$1&96 & $\pm$1&68 &
$\pm$1&33 & $\pm$1&00 & $\pm$0&75 & $\pm$0&55 & $\pm$0&36 & $\pm$0&18 &
$\pm$0&07 \\ \hline
0.00 & 1&00 & 0&94 & 0&95 & 0&91 & 0&78 & 0&78 & 0&87 & 0&84 & 0&83 & 0&70 & 0&71 & 0&60  \\
0.10 & 0&94 & 1&00 & 0&94 & 0&90 & 0&66 & 0&68 & 0&91 & 0&78 & 0&81 & 0&61 & 0&66 & 0&62  \\
0.20 & 0&95 & 0&94 & 1&00 & 0&97 & 0&83 & 0&84 & 0&90 & 0&82 & 0&79 & 0&76 & 0&70 & 0&53  \\
0.30 & 0&91 & 0&90 & 0&97 & 1&00 & 0&81 & 0&83 & 0&93 & 0&73 & 0&70 & 0&76 & 0&60 & 0&41  \\
0.40 & 0&78 & 0&66 & 0&83 & 0&81 & 1&00 & 0&99 & 0&67 & 0&81 & 0&70 & 0&89 & 0&72 & 0&37  \\
0.50 & 0&78 & 0&68 & 0&84 & 0&83 & 0&99 & 1&00 & 0&73 & 0&79 & 0&68 & 0&91 & 0&69 & 0&34  \\
0.60 & 0&87 & 0&91 & 0&90 & 0&93 & 0&67 & 0&73 & 1&00 & 0&71 & 0&72 & 0&69 & 0&60 & 0&49  \\
0.70 & 0&84 & 0&78 & 0&82 & 0&73 & 0&81 & 0&79 & 0&71 & 1&00 & 0&97 & 0&77 & 0&91 & 0&78  \\
0.80 & 0&83 & 0&81 & 0&79 & 0&70 & 0&70 & 0&68 & 0&72 & 0&97 & 1&00 & 0&72 & 0&93 & 0&87  \\
1.00 & 0&70 & 0&61 & 0&76 & 0&76 & 0&89 & 0&91 & 0&69 & 0&77 & 0&72 & 1&00 & 0&80 & 0&43  \\
1.20 & 0&71 & 0&66 & 0&70 & 0&60 & 0&72 & 0&69 & 0&60 & 0&91 & 0&93 & 0&80 & 1&00 & 0&87  \\
1.50-2.00 & 0&60 & 0&62 & 0&53 & 0&41 & 0&37 & 0&34 & 0&49 & 0&78 & 0&87 & 0&43 & 0&87 & 1&00
\end{tabular}
\end{table}
\end{turnpage}

\begin{turnpage}
\begin{table}[h!]
\caption{The flux-averaged observable \ccpiz differential cross section
 in $E_\mu$ over the flux range $E_\nu\in(0.5-2.0\;$GeV$)$ in
bins of $E_\mu-m_\mu$ (Fig.~\ref{fig:dxsecdemu}).  Tabulated are the central value  
(CV), the total systematic error, the statistical error, and the correlation
matrix for the systematic error.  The
correlations for the statistical errors are small and not tabulated.  The CV,
total systematic error, and statistical errors are multiplied by
$10^{39}$. The bin boundaries are tabulated as the low edge of the bin with
the last bin giving both the low and high edge. }
\label{tab:dxsecdemu}
\begin{tabular}{c|r@{.}lr@{.}lr@{.}lr@{.}lr@{.}lr@{.}lr@{.}lr@{.}lr@{.}lr@{.}lr@{.}lr@{.}lr@{.}l}
Bin edge [GeV] & 0&00 & 0&05 & 0&10 & 0&15 & 0&20 & 0&28 & 0&35 & 0&42 & 0&50 & 0&60 & 0&70 & 0&80 & \multicolumn{2}{c}{1.00-1.20} \\ \hline\hline
CV [$\times10^{-39}\;$cm$^2$/GeV] & 10&31 & 15&60 & 17&00 & 16&58 & 15&33 & 13&57 & 11&38 & 9&76 & 6&97 & 5&82 & 4&03 & 2&55 & 1&38 \\
Stat. & $\pm$0&96 & $\pm$1&03 & $\pm$0&75 & $\pm$0&63 & $\pm$0&48 & $\pm$0&47 & $\pm$0&45 & $\pm$0&43 & $\pm$0&32 & $\pm$0&32 & $\pm$0&28 & $\pm$0&19 & $\pm$0&21 \\
Total Syst. & $\pm$1&83 & $\pm$2&52 & $\pm$2&73 & $\pm$3&00 & $\pm$2&52 & $\pm$2&21 & $\pm$1&91 & $\pm$1&60 & $\pm$1&38 & $\pm$1&12 & $\pm$0&76 & $\pm$0&58 & $\pm$0&36 \\ \hline
0.00 & 1&00 & 0&50 & 0&75 & 0&70 & 0&66 & 0&82 & 0&75 & 0&65 & 0&76 & 0&89 & 0&62 & 0&31 & 0&34  \\
0.05 & 0&50 & 1&00 & 0&87 & 0&83 & 0&88 & 0&78 & 0&80 & 0&85 & 0&67 & 0&57 & 0&86 & 0&78 & 0&56  \\
0.10 & 0&75 & 0&87 & 1&00 & 0&93 & 0&91 & 0&92 & 0&92 & 0&90 & 0&85 & 0&82 & 0&87 & 0&67 & 0&51  \\
0.15 & 0&70 & 0&83 & 0&93 & 1&00 & 0&83 & 0&90 & 0&87 & 0&90 & 0&91 & 0&81 & 0&90 & 0&61 & 0&42  \\
0.20 & 0&66 & 0&88 & 0&91 & 0&83 & 1&00 & 0&91 & 0&93 & 0&90 & 0&75 & 0&77 & 0&87 & 0&84 & 0&70  \\
0.28 & 0&82 & 0&78 & 0&92 & 0&90 & 0&91 & 1&00 & 0&95 & 0&91 & 0&91 & 0&92 & 0&88 & 0&66 & 0&58  \\
0.35 & 0&75 & 0&80 & 0&92 & 0&87 & 0&93 & 0&95 & 1&00 & 0&93 & 0&85 & 0&87 & 0&88 & 0&75 & 0&65  \\
0.42 & 0&65 & 0&85 & 0&90 & 0&90 & 0&90 & 0&91 & 0&93 & 1&00 & 0&89 & 0&78 & 0&92 & 0&79 & 0&57  \\
0.50 & 0&76 & 0&67 & 0&85 & 0&91 & 0&75 & 0&91 & 0&85 & 0&89 & 1&00 & 0&89 & 0&84 & 0&53 & 0&38  \\
0.60 & 0&89 & 0&57 & 0&82 & 0&81 & 0&77 & 0&92 & 0&87 & 0&78 & 0&89 & 1&00 & 0&78 & 0&47 & 0&45  \\
0.70 & 0&62 & 0&86 & 0&87 & 0&90 & 0&87 & 0&88 & 0&88 & 0&92 & 0&84 & 0&78 & 1&00 & 0&77 & 0&55  \\
0.80 & 0&31 & 0&78 & 0&67 & 0&61 & 0&84 & 0&66 & 0&75 & 0&79 & 0&53 & 0&47 & 0&77 & 1&00 & 0&72  \\
1.00-1.20 & 0&34 & 0&56 & 0&51 & 0&42 & 0&70 & 0&58 & 0&65 & 0&57 & 0&38 & 0&45 & 0&55 & 0&72 & 1&00
\end{tabular}
\end{table}
\end{turnpage}

\begin{turnpage}
\begin{table}[h!]
\caption{The flux-averaged observable \ccpiz differential cross section
 in $\cos\theta_\mu$ over the flux range
$E_\nu\in(0.5-2.0\;$GeV$)$ (Fig.~\ref{fig:dxsecdcosmu}).  Tabulated are the central value  
(CV), the total systematic error, the statistical error, and the correlation
matrix for the systematic error.  The
correlations for the statistical errors are small and not tabulated.  The CV,
total systematic error, and statistical errors are multiplied by
$10^{39}$. The bin boundaries are tabulated as the low edge of the bin with
the last bin giving both the low and high edge.}
\label{tab:dxsecdcosmu}
\begin{tabular}{c|r@{.}lr@{.}lr@{.}lr@{.}lr@{.}lr@{.}lr@{.}lr@{.}lr@{.}l}
Bin edge& -1&00 & -0&60 & -0&30 & 0&00 & 0&20 & 0&40 & 0&60 & 0&80 & \multicolumn{2}{c}{0.90-1.00} \\ \hline\hline
CV [$\times10^{-39}\;$cm$^2$] & 1&01 & 1&55 & 2&46 & 3&60 & 4&65 & 6&58 & 10&32 & 13&11 & 11&51 \\
Stat. & $\pm$0&08 & $\pm$0&11 & $\pm$0&13 & $\pm$0&18 & $\pm$0&19 & $\pm$0&23 & $\pm$0&28 & $\pm$0&47 & $\pm$0&46 \\
Total Syst. & $\pm$0&25 & $\pm$0&32 & $\pm$0&40 & $\pm$0&54 & $\pm$0&76 & $\pm$1&22 & $\pm$1&59 & $\pm$2&29 & $\pm$2&92 \\ \hline
-1.00 & 1&00 & 0&61 & 0&67 & 0&45 & 0&58 & 0&86 & 0&34 & 0&66 & 0&53  \\
-0.60 & 0&61 & 1&00 & 0&84 & 0&63 & 0&87 & 0&77 & 0&70 & 0&84 & 0&71  \\
-0.30 & 0&67 & 0&84 & 1&00 & 0&78 & 0&90 & 0&83 & 0&80 & 0&92 & 0&83  \\
0.00 & 0&45 & 0&63 & 0&78 & 1&00 & 0&76 & 0&64 & 0&90 & 0&80 & 0&87  \\
0.20 & 0&58 & 0&87 & 0&90 & 0&76 & 1&00 & 0&79 & 0&84 & 0&92 & 0&84  \\
0.40 & 0&86 & 0&77 & 0&83 & 0&64 & 0&79 & 1&00 & 0&60 & 0&87 & 0&76  \\
0.60 & 0&34 & 0&70 & 0&80 & 0&90 & 0&84 & 0&60 & 1&00 & 0&84 & 0&92  \\
0.80 & 0&66 & 0&84 & 0&92 & 0&80 & 0&92 & 0&87 & 0&84 & 1&00 & 0&92  \\
0.90-1.00 & 0&53 & 0&71 & 0&83 & 0&87 & 0&84 & 0&76 & 0&92 & 0&92 & 1&00
\end{tabular}
\end{table}
\end{turnpage}

\begin{turnpage}
\begin{table}[h!]
\caption{The flux-averaged observable \ccpiz differential cross section
 in $|\mathbf{p_{\pi^0}}|$ over the flux range
$E_\nu\in(0.5-2.0\;$GeV$)$ (Fig.~\ref{fig:dxsecdppi}). Tabulated are the central value  
(CV), the total systematic error, the statistical error, and the correlation
matrix for the systematic error.  The
correlations for the statistical errors are small and not tabulated.  The CV,
total systematic error, and statistical errors are multiplied by
$10^{39}$. The bin boundaries are tabulated as the low edge of the bin with
the last bin giving both the low and high edge.}
\label{tab:dxsecdppi}
\begin{tabular}{c|r@{.}lr@{.}lr@{.}lr@{.}lr@{.}lr@{.}lr@{.}lr@{.}lr@{.}lr@{.}lr@{.}l}
Bin edge [GeV/$c$] & 0&00 & 0&10 & 0&15 & 0&20 & 0&25 & 0&30 & 0&40 & 0&50 & 0&60 & 0&80 & \multicolumn{2}{c}{1.00-1.40} \\ \hline\hline
CV [$\times10^{39}\;$cm$^2$/GeV/$c$] & 4&92 & 26&65 & 32&90 & 28&99 & 19&02 & 13&65 & 7&41 & 4&27 & 1&90 & 0&87 & 0&19 \\
Stat. & $\pm$0&37 & $\pm$1&10 & $\pm$1&05 & $\pm$0&87 & $\pm$0&61 & $\pm$0&37 & $\pm$0&31 & $\pm$0&33 & $\pm$0&25 & $\pm$0&35 & $\pm$0&21 \\
Total Syst. & $\pm$1&34 & $\pm$4&94 & $\pm$5&00 & $\pm$4&31 & $\pm$3&09 & $\pm$2&49 & $\pm$2&01 & $\pm$1&14 & $\pm$0&63 & $\pm$0&40 & $\pm$0&79 \\ \hline
0.00 & 1&00 & 0&74 & 0&83 & 0&52 & 0&82 & 0&79 & 0&47 & 0&25 & 0&26 & 0&44 & 0&02  \\
0.10 & 0&74 & 1&00 & 0&92 & 0&74 & 0&90 & 0&70 & 0&30 & 0&51 & 0&58 & 0&26 & -0&41  \\
0.15 & 0&83 & 0&92 & 1&00 & 0&85 & 0&97 & 0&88 & 0&56 & 0&59 & 0&57 & 0&36 & -0&14  \\
0.20 & 0&52 & 0&74 & 0&85 & 1&00 & 0&82 & 0&84 & 0&73 & 0&84 & 0&72 & 0&26 & 0&05  \\
0.25 & 0&82 & 0&90 & 0&97 & 0&82 & 1&00 & 0&90 & 0&55 & 0&61 & 0&63 & 0&38 & -0&16  \\
0.30 & 0&79 & 0&70 & 0&88 & 0&84 & 0&90 & 1&00 & 0&84 & 0&69 & 0&57 & 0&48 & 0&24  \\
0.40 & 0&47 & 0&30 & 0&56 & 0&73 & 0&55 & 0&84 & 1&00 & 0&72 & 0&46 & 0&44 & 0&63  \\
0.50 & 0&25 & 0&51 & 0&59 & 0&84 & 0&61 & 0&69 & 0&72 & 1&00 & 0&89 & 0&35 & 0&08  \\
0.60 & 0&26 & 0&58 & 0&57 & 0&72 & 0&63 & 0&57 & 0&46 & 0&89 & 1&00 & 0&39 & -0&19  \\
0.80 & 0&44 & 0&26 & 0&36 & 0&26 & 0&38 & 0&48 & 0&44 & 0&35 & 0&39 & 1&00 & 0&24  \\
1.00-1.40 & 0&02 & -0&41 & -0&14 & 0&05 & -0&16 & 0&24 & 0&63 & 0&08 & -0&19 & 0&24 & 1&00
\end{tabular}
\end{table}
\end{turnpage}

\begin{turnpage}
\begin{table}[h!]
\caption{The flux-averaged observable \ccpiz differential cross section
 in $\cos\theta_{\pi^0}$ over the flux range
$E_\nu\in(0.5-2.0\;$GeV$)$ (Fig.~\ref{fig:dxsecdcospi}). Tabulated are the central value  
(CV), the total systematic error, the statistical error, and the correlation
matrix for the systematic error.  The
correlations for the statistical errors are small and not tabulated.  The CV,
total systematic error, and statistical errors are multiplied by
$10^{39}$. The bin boundaries are tabulated as the low edge of the bin with
the last bin giving both the low and high edge.}
\label{tab:dxsecdcospi}
\begin{tabular}{c|r@{.}lr@{.}lr@{.}lr@{.}lr@{.}lr@{.}lr@{.}lr@{.}lr@{.}lr@{.}lr@{.}lr@{.}lr@{.}lr@{.}lr@{.}l}
Bin edge & -1&00 & -0&70 & -0&40 & -0&20 & -0&10 & 0&00 & 0&10 & 0&20 & 0&30 &
0&40 & 0&50 & 0&60 & 0&70 & 0&80 & \multicolumn{2}{c}{0.90-1.00} \\ \hline\hline
CV [$\times10^{-39}\,$cm$^2$] & 2&34 & 2&45 & 2&83 & 2&76 & 2&77 & 3&06 & 3&55 & 4&01 & 4&57 & 5&47 & 6&44 & 7&66 & 9&14 & 11&36 & 14&59 \\
Stat. & $\pm$0&16 & $\pm$0&12 & $\pm$0&14 & $\pm$0&18 & $\pm$0&17 & $\pm$0&18 & $\pm$0&19 & $\pm$0&21 & $\pm$0&22 & $\pm$0&26 & $\pm$0&29 & $\pm$0&33 & $\pm$0&39 & $\pm$0&49 & $\pm$0&66 \\ 
Total Syst. & $\pm$0&44 & $\pm$0&45 & $\pm$0&50 & $\pm$0&46 & $\pm$0&49 &
$\pm$0&59 & $\pm$0&60 & $\pm$0&69 & $\pm$0&76 & $\pm$0&95 & $\pm$1&05 &
$\pm$1&26 & $\pm$1&48 & $\pm$2&08 & $\pm$2&83 \\ \hline
-1.00 & 1&00 & 0&87 & 0&78 & 0&78 & 0&84 & 0&62 & 0&85 & 0&84 & 0&78 & 0&85 & 0&81 & 0&86 & 0&80 & 0&63 & 0&81  \\
-0.70 & 0&87 & 1&00 & 0&70 & 0&78 & 0&87 & 0&63 & 0&91 & 0&81 & 0&71 & 0&92 & 0&86 & 0&88 & 0&80 & 0&60 & 0&88  \\
-0.40 & 0&78 & 0&70 & 1&00 & 0&67 & 0&69 & 0&48 & 0&76 & 0&93 & 0&86 & 0&71 & 0&70 & 0&80 & 0&80 & 0&60 & 0&68  \\
-0.20 & 0&78 & 0&78 & 0&67 & 1&00 & 0&96 & 0&92 & 0&90 & 0&75 & 0&82 & 0&84 & 0&93 & 0&89 & 0&90 & 0&90 & 0&84  \\
-0.10 & 0&84 & 0&87 & 0&69 & 0&96 & 1&00 & 0&88 & 0&95 & 0&80 & 0&82 & 0&91 & 0&94 & 0&91 & 0&89 & 0&82 & 0&90  \\
0.00 & 0&62 & 0&63 & 0&48 & 0&92 & 0&88 & 1&00 & 0&80 & 0&59 & 0&75 & 0&72 & 0&87 & 0&77 & 0&81 & 0&90 & 0&75  \\
0.10 & 0&85 & 0&91 & 0&76 & 0&90 & 0&95 & 0&80 & 1&00 & 0&89 & 0&84 & 0&95 & 0&93 & 0&94 & 0&91 & 0&77 & 0&93  \\
0.20 & 0&84 & 0&81 & 0&93 & 0&75 & 0&80 & 0&59 & 0&89 & 1&00 & 0&92 & 0&85 & 0&83 & 0&88 & 0&87 & 0&68 & 0&82  \\
0.30 & 0&78 & 0&71 & 0&86 & 0&82 & 0&82 & 0&75 & 0&84 & 0&92 & 1&00 & 0&80 & 0&86 & 0&87 & 0&90 & 0&83 & 0&80  \\
0.40 & 0&85 & 0&92 & 0&71 & 0&84 & 0&91 & 0&72 & 0&95 & 0&85 & 0&80 & 1&00 & 0&92 & 0&92 & 0&87 & 0&71 & 0&95  \\
0.50 & 0&81 & 0&86 & 0&70 & 0&93 & 0&94 & 0&87 & 0&93 & 0&83 & 0&86 & 0&92 & 1&00 & 0&95 & 0&93 & 0&88 & 0&92  \\
0.60 & 0&86 & 0&88 & 0&80 & 0&89 & 0&91 & 0&77 & 0&94 & 0&88 & 0&87 & 0&92 & 0&95 & 1&00 & 0&95 & 0&81 & 0&93  \\
0.70 & 0&80 & 0&80 & 0&80 & 0&90 & 0&89 & 0&81 & 0&91 & 0&87 & 0&90 & 0&87 & 0&93 & 0&95 & 1&00 & 0&88 & 0&90  \\
0.80 & 0&63 & 0&60 & 0&60 & 0&90 & 0&82 & 0&90 & 0&77 & 0&68 & 0&83 & 0&71 & 0&88 & 0&81 & 0&88 & 1&00 & 0&77  \\
0.90-1.00 & 0&81 & 0&88 & 0&68 & 0&84 & 0&90 & 0&75 & 0&93 & 0&82 & 0&80 & 0&95 & 0&92 & 0&93 & 0&90 & 0&77 & 1&00
\end{tabular}
\end{table}
\end{turnpage}

\bibliography{ccpi0_prd}

\end{document}